\documentclass[
tightenlines,
twocolumn,
prc,
a4paper,
nofootinbib,
superscriptaddress,
floatfix,
preprintnumbers,
flushbottom,
aps
]{revtex4-1}

\usepackage{url}
\usepackage[%
        hyperindex,breaklinks,
	pdfstartview=FitH,
  pdfpagemode=UseNone
	]{hyperref}

\usepackage[
   centertags, 
   sumlimits,  
   intlimits,  
   namelimits, 
]{amsmath} %
\usepackage{amssymb}

\makeatletter
\def\tagform@#1{\maketag@@@{\ignorespaces#1\unskip\@@italiccorr}}
\let\orgtheequation\theequation
\def\theequation{(\orgtheequation)}
\makeatother

\usepackage[all]{hypcap} 
\renewcommand{\BibitemShut}[1]{}

\usepackage[english]{babel} 

\usepackage[%
]{graphicx}

\usepackage{xspace}
\newcommand{\Ref}{Ref.}
\newcommand{\Refs}{Refs.}

\newcommand{\jpsi}{$J/\psi$\xspace}
\newcommand{\unit}[1]{\, {\rm #1}}
\newcommand{\breakeq}{\nonumber\\}

\newcommand{\arr}{\rightarrow}
\newcommand{\sqrs}{\sqrt{s}}
\def\l{\left}
\def\r{\right}
\def\d{{\rm d}}

\newcommand{\cm}{center-of-mass\xspace}

\newcommand{\vt}{$v_2$\xspace}
\newcommand{\raa}{$R_{AA}$\xspace}

\begin{document}
\title{Elastic and radiative heavy quark interactions in ultra-relativistic heavy-ion collisions}


\author{Jan Uphoff}
\email[E-mail: ]{uphoff@th.physik.uni-frankfurt.de}

\author{Oliver Fochler}
\affiliation{Institut f\"ur Theoretische Physik, Goethe-Universit\"at Frankfurt, Max-von-Laue-Str.\ 1, 
D-60438 Frankfurt am Main, Germany}

\author{Zhe Xu}
\affiliation{Department of Physics, Tsinghua University, Beijing 100084, China}
\affiliation{Collaborative Innovation Center of Quantum Matter, Beijing, China}

\author{Carsten Greiner}
\affiliation{Institut f\"ur Theoretische Physik, Goethe-Universit\"at Frankfurt, Max-von-Laue-Str.\ 1, 
D-60438 Frankfurt am Main, Germany}

\date{\today}

\begin{abstract}
Elastic and radiative heavy quark interactions with light partons are studied with the partonic transport description BAMPS (\emph{Boltzmann Approach to MultiParton Scatterings}). After calculating the cross section of radiative processes for finite masses in the improved Gunion-Bertsch approximation and verifying this calculation by comparing to the exact result, we study elastic and radiative heavy quark energy loss in a static medium of quarks and gluons. Furthermore, the full 3+1D space-time evolution of gluons, light quarks, and heavy quarks in ultra-relativistic heavy-ion collisions at the BNL Relativistic Heavy-Ion Collider (RHIC) and the CERN Large Hadron Collider (LHC) are calculated with BAMPS including elastic and radiative heavy flavor interactions. Treating light and heavy particles on the same footing in the same framework, we find that the experimentally measured nuclear modification factor of charged hadrons and $D$ mesons at the LHC can be simultaneously described. In addition, we calculate the heavy flavor evolution with an improved screening procedure from hard-thermal-loop calculations and confront the results with experimental data of the nuclear modification factor and the elliptic flow of heavy flavor particles at RHIC and LHC.
\end{abstract}


\maketitle

\section{Introduction}

Ultra-relativistic heavy-ion collisions are a unique tool to study the properties of the interactions between quarks and gluons. Due to the large energy density created in such collisions quarks and gluons are not confined in protons and neutrons any more, but are the relevant degrees of freedom forming the quark-gluon plasma (QGP). Many experimental observations at the BNL Relativistic Heavy-Ion Collider (RHIC) and the CERN Large Hadron Collider (LHC) indicate that such a plasma is indeed produced in heavy-ion collisions at these colliders (for reviews see, for instance, Ref.~\cite{Jacak:2012dx,Muller:2012zq}). Important observations include collective effects, most prominently the elliptic flow in semi-central collisions, and the quenching of jets through the nuclear modification factor in central collisions. 

Heavy quarks are a particularly helpful probe for studying the properties of the medium. Their creation time is well defined since they are either produced in initial hard parton scatterings or due to their large mass in the early stage of the QGP evolution, where the energy density is still large \cite{Uphoff:2010sh}. While traversing the created medium they lose energy and participate in the collective flow. Therefore, heavy flavor observables can reveal insights in the properties of the medium itself as well as the interaction strength between heavy and light particles. Furthermore, the quark flavor (for heavy quarks in heavy-ion collisions charm and bottom is most relevant) is conserved during hadronization such that the associated heavy mesons ($D$ and $B$ mesons coming from charm and bottom, respectively) or their decay products are in principle uniquely related to the heavy quarks.

The existing literature on heavy flavor interaction mechanisms can be sorted into three categories: 1) pQCD inspired interactions \cite{Moore:2004tg, Mustafa:2004dr, Zhang:2005ni, Molnar:2006ci, Armesto:2005mz, Djordjevic:2005db,Wicks:2005gt,Djordjevic:2011tm,Djordjevic:2013pba, Gossiaux:2008jv,Gossiaux:2009mk,Gossiaux:2012ya,Nahrgang:2013saa, Sharma:2009hn, Uphoff:2011ad,Uphoff:2012gb, Alberico:2011zy,Alberico:2013bza, Buzzatti:2011vt,Buzzatti:2012pe, Young:2011ug,  Mazumder:2011nj,Mazumder:2013oaa, Cao:2011et,Cao:2012au,Cao:2013ita, Meistrenko:2012ju, Abir:2012pu, Renk:2013xaa}, 
2) resonance scatterings \cite{Adil:2006ra, vanHees:2005wb,vanHees:2007me, He:2011qa,He:2012df, Lang:2012yf,Lang:2012cx}, and 
3) strong coupling interactions calculated from AdS/CFT correspondence \cite{Horowitz:2007su,Horowitz:2012cf, Chesler:2012pw,Chesler:2013cqa}. 
The pQCD inspired models include either elastic \cite{Moore:2004tg, Zhang:2005ni, Molnar:2006ci, Gossiaux:2008jv,Gossiaux:2009mk, Uphoff:2011ad,Uphoff:2012gb, Alberico:2011zy,Alberico:2013bza, Young:2011ug, Meistrenko:2012ju} or radiative \cite{Abir:2012pu} processes or both \cite{Mustafa:2004dr, Armesto:2005mz, Djordjevic:2005db,Wicks:2005gt,Djordjevic:2011tm,Djordjevic:2013pba, Sharma:2009hn, Buzzatti:2011vt,Buzzatti:2012pe, Mazumder:2011nj,Mazumder:2013oaa, Cao:2011et,Cao:2012au,Cao:2013ita, Gossiaux:2012ya,Nahrgang:2013saa, Renk:2013xaa}.

In this paper we study heavy quark interactions based on pQCD cross sections with the partonic transport description BAMPS via coupled relativistic Boltzmann equations. Previous heavy flavor BAMPS calculations \cite{Uphoff:2011ad,Uphoff:2012gb,Uphoff:2013rka} included only elastic interactions with the rest of the QGP. In this paper we introduce the implementation of the radiative heavy flavor processes in BAMPS, compare them to elastic interactions in a static medium, and confront the nuclear modification factor and elliptic flow of various heavy flavor particles with experimental measurements at RHIC and LHC.

The paper is organized as follows. After a short introduction to BAMPS in the next section, we generalize the improved Gunion-Bertsch calculation for the radiative matrix element from Ref.~\cite{Fochler:2013epa} to finite masses in Sec.~\ref{sec:gb_cs}. The full calculation is shown in the Appendix.
In Secs.~\ref{sec:dead_cone_effect} and~\ref{sec:comparison_gb_exact} we show that the dead cone effect is present in our expression for heavy flavor radiation and compare the improved Gunion-Bertsch approximation to the exact calculation, respectively. Furthermore, the LPM implementation is described in detail in Secs.~\ref{sec:lpm} and~\ref{sec:phase_space_lpm}. In Sec.~\ref{sec:eloss_radiative} the elastic and radiative energy loss of heavy and light partons in a static medium is calculated and put into context. Full heavy-ion collisions are considered in Sec.~\ref{sec:results_open_hf_radiative}. First we compare the suppression of heavy and light particles with standard Debye screening to experimental data. In Sec.~\ref{sec:23_hf_kappa} heavy flavor observables are calculated for an improved screening procedure and confronted to experimental data at RHIC and LHC. Finally, we conclude with Sec.~\ref{sec:conclusions}.
Many results of this paper are based on Ref.~\cite{uphoff_phd}. We refer to this thesis also for additional information and in-depth analysis of heavy flavor processes.

\section{The parton cascade BAMPS}

The partonic transport model BAMPS (\emph{Boltzmann Approach to MultiParton Scatterings}) \cite{Xu:2004mz,Xu:2007aa,Uphoff:2014cba} has been successfully applied to study the partonic evolution in ultra-relativistic heavy-ion collisions by solving the Boltzmann transport equation,
\begin{equation}
\label{boltzmann}
\left ( \frac{\partial}{\partial t} + \frac{{\mathbf p}}{E}
\frac{\partial}{\partial {\mathbf r}} \right )\, 
f_i({\mathbf r}, {\mathbf p}, t) = {\cal C}_i^{2\rightarrow 2} + {\cal C}_i^{2\leftrightarrow 3}+ \ldots  \ ,
\end{equation}
for on-shell partons. The interaction probabilities are calculated from pQCD cross sections for elastic and inelastic partonic processes. Recently, it has been shown \cite{Uphoff:2014cba} that the explicit consideration of the running of the coupling together with the improved Gunion-Bertsch cross section \cite{Fochler:2013epa} and an effective description of the LPM effect can describe the experimentally measured data of the nuclear modification factor for neutral pions at RHIC and charged hadrons at LHC. Employing the same cross sections within the same model also a sizeable elliptic flow on the partonic level is observed, which is due to a small shear viscosity to entropy density ratio \cite{Uphoff:2014cba}.

Previous BAMPS studies on the heavy flavor sector \cite{Uphoff:2011ad,Uphoff:2012gb,Uphoff:2013rka} included only elastic heavy quark interactions with light partons from the rest of the medium. The cross sections of such elastic interactions are calculated in leading order pQCD and explicitly take into account the running of the coupling. Furthermore, the screening mass $\mu$ to regularize infrared divergences is calculated by matching hard thermal loop results for the energy loss of heavy quarks in a static medium \cite{Gossiaux:2008jv,Peshier:2008bg,Uphoff:2011ad}. To this end, the screening mass is proportional to the Debye mass~$m_D$, $\mu^2 = \kappa_t m_D^2$ with $\kappa_t = 1/(2e) \approx 0.2$. 

The initial heavy flavor distribution is obtained from MC@NLO \cite{Frixione:2002ik,Frixione:2003ei} in next-to-leading order, which agrees well with experimental data of heavy flavor particles in proton-proton collisions \cite{Uphoff:2012gb}. The initial light parton distribution is obtained from \textsc{pythia} \cite{Sjostrand:2006za,Sjostrand:2007gs} as described in Ref.~\cite{Uphoff:2010sh}. The spatial distribution of all particles is sampled according to the Glauber model \cite{Xu:2004mz,Uphoff:2010sh}.
Details on the heavy flavor implementation in BAMPS can be found in Refs.~\cite{Uphoff:2010sh,Uphoff:2011ad,Uphoff:2012gb}.

In the following we introduce the implementation of radiative bremsstrahlung processes off a heavy quark. These $2 \rightarrow 3$ interactions are calculated in the improved Gunion-Bertsch approximation, which is outlined in the next section. After we show in Sec.~\ref{sec:dead_cone_effect} that the dead cone effect is explicitly present in our result, we compare in Sec.~\ref{sec:comparison_gb_exact} the Gunion-Bertsch approximation to the exact result. In Sec.~\ref{sec:lpm} we introduce the implemented description of the LPM effect, which is analogous to the implementation of light partons in BAMPS, but explicitly takes the finite mass of heavy quarks into account. Thereafter, some kinematic consequences of the LPM implementation are highlighted in Sec.~\ref{sec:phase_space_lpm}.


\subsection{Gunion-Bertsch matrix element for heavy quarks}
\label{sec:gb_cs}

This approximation gives a comparatively simple expression for the gluon radiation amplitude in terms of the transverse momentum of the radiated gluon $k_{\perp}$ and the transverse exchanged momentum $q_{\perp}$. 

As outlined in the Appendix, 
the total squared matrix element for the process $qQ\arr qQg$ summed and averaged over spin, polarization, and color states reads
\begin{multline}
\label{gb_qQg_matrix_element_xOrg}
	{\l|\overline{\mathcal{M}}_{qQ \rightarrow qQg}\r|}^2 
	=12 g^2  \,
  \l|\overline{\mathcal{M}}_{qQ\arr qQ}\r|^2 \, (1-x)^2
\\ \times
	\l[ \frac{ {\bf k}_\perp}{k_\perp^2+x^2M^2} +  \frac{ {\bf q}_\perp - {\bf k}_\perp}{({\bf q}_\perp - {\bf k}_\perp)^2+x^2M^2} \r]^2 \ .
\end{multline}
$\l|\overline{\mathcal{M}}_{qQ\arr qQ}\r|^2$ denotes the averaged squared matrix element for $qQ\arr qQ$. This result explicitly demonstrates that in the high-energy limit the matrix element of the $2\arr 3$ process factorizes into an elastic $2\arr 2$ part and a gluon emission amplitude.
For the massless case, $M=0$, Eq.~\eqref{gb_qQg_matrix_element_xOrg} reduces to the improved GB result from Ref.~\cite{Fochler:2013epa}.

As outlined in Ref.~\cite{Fochler:2013epa} the factor $(1-x)^{2}$ in Eq.~\eqref{gb_qQg_matrix_element_xOrg} leads to a sizeable suppression of the amplitude at forward rapidity, where $x > k_{\perp}/\sqrs$, which is immediately evident from Eq.~\eqref{x_def_y}. This factor is not present in the final result of Gunion and Bertsch, while it is included in intermediate steps of their calculation. In Ref.~\cite{Fochler:2013epa} we also argued that the calculation in $A^+ = 0$ gauge with the employed approximations is not valid at backward rapidity, which also holds for the heavy quark calculation. To this end, one can do the same calculation in $A^- = 0$ gauge and symmetrize the result.
 Setting
\begin{align}
\label{xdash_def_y}
x' = \frac{k_\perp}{\sqrs} {\rm e}^{-y} \ ,
\end{align} 
the final result for the averaged squared matrix element in $A^- = 0$ gauge reads
\begin{multline}
\label{gb_qQg_matrix_element_xdash}
	{\l|\overline{\mathcal{M}}_{qQ \rightarrow qQg}\r|}^2 =12 g^2  \,
  \l|\overline{\mathcal{M}}_{qQ\arr qQ}\r|^2 \, (1-x')^2
\\ \times
	\l[ \frac{ {\bf k}_\perp}{k_\perp^2} +  \frac{ {\bf q}_\perp - {\bf k}_\perp}{({\bf q}_\perp - {\bf k}_\perp)^2} \r]^2 \ .
\end{multline} 
Note that no mass terms are present in the denominators of the bracket since the gluon is now emitted off the light quark line. 
Apart from this, Eq.~\eqref{gb_qQg_matrix_element_xdash} is simply the result obtained from the $A^+ = 0$ gauge, cf.~\eqref{gb_qQg_matrix_element_xOrg}, with $x$ being replaced by $x'$. Since both results are also valid at mid-rapidity it is self-evident to combine Eqs.~\eqref{gb_qQg_matrix_element_xOrg} and \eqref{gb_qQg_matrix_element_xdash} to
\begin{widetext}
\begin{align}
\label{gb_qQg_matrix_element_improved}
	{\l|\overline{\mathcal{M}}_{qQ \rightarrow qQg}\r|}^2 &=12 g^2  \,
  \l|\overline{\mathcal{M}}_{qQ\arr qQ}\r|^2 \, 
  \l[ \Theta(y) (1-x)^2
	\l[ \frac{ {\bf k}_\perp}{k_\perp^2+x^2M^2} +  \frac{ {\bf q}_\perp - {\bf k}_\perp}{({\bf q}_\perp - {\bf k}_\perp)^2+x^2M^2} \r]^2 
\r. 
\breakeq 
  & \qquad \qquad \qquad \qquad \qquad \qquad \l. 
+ \Theta(-y) (1-x')^2
	\l[ \frac{ {\bf k}_\perp}{k_\perp^2} +  \frac{ {\bf q}_\perp - {\bf k}_\perp}{({\bf q}_\perp - {\bf k}_\perp)^2} \r]^2 \r]\breakeq 
		&\simeq  12 g^2  \,
  \l|\overline{\mathcal{M}}_{qQ\arr qQ}\r|^2 \,  (1-\bar{x})^2
	\l[ \frac{ {\bf k}_\perp}{k_\perp^2+x^2M^2} +  \frac{ {\bf q}_\perp - {\bf k}_\perp}{({\bf q}_\perp - {\bf k}_\perp)^2+x^2M^2} \r]^2 \ ,
\end{align}
\end{widetext}
where we have defined
\begin{align}
\label{xbar_def_y}
\bar{x} = \frac{k_\perp}{\sqrs} {\rm e}^{|y|}\ .
\end{align}
Going from the first to the second line we added the terms $x^2M^2$ also in the denominators of the backward rapidity terms proportional to $\Theta(-y)$. This is valid within the approximations since $x$ (not $x'$ or $\bar x$) is small at backward rapidity according to its definition~\eqref{x_def_y}.
In the following we refer to this result as the improved GB matrix element for heavy quarks since it is not only valid at mid-rapidity, but also at forward and backward rapidity. In Sec.~\ref{sec:comparison_gb_exact} we compare this result to the exact matrix element without any approximations.

As a note, the idea of decomposing the phase space in forward and backward rapidities was also used in \Ref~\cite{guiho_phd} for the inelastic scattering of a light and heavy quark in the GB approximation, although the  $(1-\bar x)^2$ factor was neglected.

In Ref.~\cite{Abir:2011jb} the matrix element for the process $q+Q \arr q+Q+g$ has been calculated in Feynman gauge with slightly different approximations. We checked that this calculation and the calculation in the present paper agree within their approximations, which is an independent check that our calculations are correct.

The observed factorization of the GB matrix element into an elastic part and a gluon emission amplitude in the high-energy limit, cf.\ for instance Eq.~\eqref{gb_qQg_matrix_element_improved}, immediately implies that the GB calculation is also valid for other $2 \arr 3$ processes, such as $Qg \arr Qgg$ or for light partons $qg \arr qgg$ or $gg \arr ggg$. In the high-energy limit of the GB approximations the specific nature of the scattering particles becomes irrelevant for the structure of the resulting matrix elements. Therefore, Eq.~\eqref{gb_qQg_matrix_element_improved} also holds for processes other than the one explicitly considered here if one takes into account the different color prefactors of the corresponding $2 \arr 2$ small angle amplitudes, 
\begin{align} 
\l|\overline{\mathcal{M}}_{gQ\arr gQ}\r|^2 \simeq \frac{9}{4} \, \l|\overline{\mathcal{M}}_{qQ\arr qQ}\r|^2 \ ,
\end{align}
and for light partons,
\begin{align} \label{eq:color_factors}
\l|\overline{\mathcal{M}}_{gg\arr gg}\r|^2 \simeq \frac{9}{4} \, \l|\overline{\mathcal{M}}_{qg\arr qg}\r|^2 \simeq \l(\frac{9}{4}\r)^2 \l|\overline{\mathcal{M}}_{qq'\arr qq'}\r|^2 \ .
\end{align}

In summary, one can split the squared matrix element of the QCD bremsstrahlung process into the contribution from the elastic scattering plus a radiative factor $P^g_M$ in the high-energy limit. For heavy quarks it reads
\begin{align}
\label{me_gb_hq}
		{\l|\overline{\mathcal{M}}_{g(q)+Q \rightarrow g(q)+Q+g}\r|}^2 =
	\l|\overline{\mathcal{M}}_{g(q)+Q \rightarrow g(q)+Q}\r|^2 \, P^g_M
\end{align}
with
\begin{multline}
\label{gb_pgm_radiation_spectrum}
	P^g_M = 12 g^2 \, (1-\bar{x})^2  \,
\\ \times
\l[ \frac{ {\bf k}_\perp}{k_\perp^2+x^2M^2} + \frac{ {\bf q}_\perp - {\bf k}_\perp}{({\bf q}_\perp - {\bf k}_\perp)^2+x^2M^2} \r]^2 \ ,
\end{multline}
which depends on the mass $M$ of the heavy quark.

The final result \eqref{me_gb_hq} is proportional to $g^6$ or $\alpha_s^3$ due to the three vertices, two associated with the $2 \arr 2$ part and one with the emitted gluon. Explicitly considering the running of the coupling,
\index{Coupling!running}
it is sensible to choose proper scales since this reduces the influence of higher orders. The relevant scale for the binary part is the momentum transfer $t$. The only scale associated with the vertex of the emitted gluon is its transverse momentum $k_\perp$. Hence, the coupling of the $2\arr 3$ matrix element goes with $\alpha_s^2(t)\, \alpha_s(k_\perp^2)$. 
Considering medium effects and the running coupling, the radiation factor from Eq.~\eqref{gb_pgm_radiation_spectrum} modifies to
\begin{multline}
\label{gb_pgm_radiation_spectrum_screened}
	P^g_M = 48 \pi \alpha_s(k_\perp^2) \, (1-\bar{x})^2  \,	
\l[ \frac{ {\bf k}_\perp}{k_\perp^2+x^2M^2} \r. 
\\ 
\l. +   \frac{ {\bf q}_\perp - {\bf k}_\perp }{ ({\bf q}_\perp - {\bf k}_\perp)^2+ m_D^2\left(\alpha_s(k_\perp^2)\right)+x^2M^2} \r]^2 \ .
\end{multline}
The coupling present in the definition of the Debye mass is also evaluated at the scale $k_\perp^2$.
The first term in the bracket does not need to be screened by a screening mass since the infrared divergence is cured by the LPM cut-off, as shown in Sec.~\ref{sec:lpm}. The second term stems from diagram~5 in Fig.~\ref{fig:feynman_diagrams_qQ_qQg} and resembles the propagator from one of the internal gluon lines. Hence, it must be screened with the gluon Debye mass.
The binary part of the matrix element reads then \cite{Combridge:1978kx}
\begin{multline}
\label{matrix_element_qQ_qQ_runAs}
{|\overline{\mathcal{M}}_{qQ \rightarrow qQ}|}^2 =
\frac{64}{9} \pi^2 \alpha_s^2(t) \\ 
\times \frac{(M^2-u)^2 + (s-M^2)^2 + 2M^2 t}{[t-\kappa_t \, m_{D}^2(\alpha_s(t))]^2} 
\end{multline}
with $\kappa_t= 1/(2e) \approx 0.2$ being the Debye mass prefactor matched to hard thermal loop calculations \cite{Gossiaux:2008jv,Peshier:2008bg,Uphoff:2011ad}. As a note, for $M=0$ and $\kappa_t=1$ these cross sections simplify to the expressions for light partons used in Ref.~\cite{Uphoff:2014cba}.


\subsection{Dead cone effect}
\label{sec:dead_cone_effect}

In this section we show that the dead cone effect, that is, the suppression of gluon emission off a heavy quark at small angles, is present in our result from Eq.~\eqref{gb_qQg_matrix_element_improved} and relate it to the suppression factor from Dokshitzer and Kharzeev \cite{Dokshitzer:2001zm}, which reads
\begin{align}
\label{dead_cone_factor}
	\mathcal{D}_{\rm DK} = \frac{ 1 }{ \l( 1 + \frac{M^2}{\theta^2 \, E^2} \r)^2 } 
	= \frac{ 1 }{ \l( 1 + \frac{\theta^2_D}{\theta^2} \r)^2 }
\end{align}
with $\theta$ being the angle between the emitted gluon and the heavy quark and $\theta_D = M / E$ the dead cone angle.

In general, the dead cone factor can be defined as the radiation spectrum of a gluon emitted off a heavy quark as given in Eq.~\eqref{gb_pgm_radiation_spectrum} divided by the  spectrum for the massless case,
\begin{align}
	\mathcal{D}_{\rm GB} = \frac{P^g_M}{P^g_0} \ .
\end{align}
In contrast to $\mathcal{D}_{\rm DK}$ an explicit dependence on $k_\perp$, $q_\perp$, and the angle between both, $\phi$, remains. For $E \gg M$, a small angle $\theta$, and $q_\perp \gg k_\perp$ the more general dead cone factor $\mathcal{D}_{\rm GB}$ simplifies to $\mathcal{D}_{\rm DK}$.

\begin{figure}
	\centering
\includegraphics[width=\linewidth]{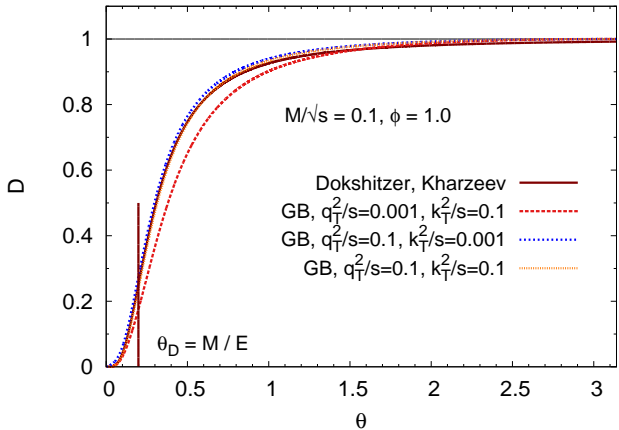}
\caption[Angle dependence of dead cone factor for $M/\sqrt{s} = 0.1$]{The dead cone factor $\mathcal{D}_{\rm GB}$ as a function of the angle $\theta$ between the emitted gluon and the heavy quark for $M/\sqrt{s} = 0.1$ and $\phi = 1.0$ (the angle between ${\bf k}_\perp$ and ${\bf q}_\perp$). The curves are for different values of $k_\perp^2/s$ and $q_\perp^2/s$. In addition, the result of \Ref~\cite{Dokshitzer:2001zm} is shown. The vertical line indicates the dead cone angle $\theta_D = M / E$.}
\label{fig:dead_cone_gb_m01}
\end{figure}

\begin{figure}
	\centering
\includegraphics[width=\linewidth]{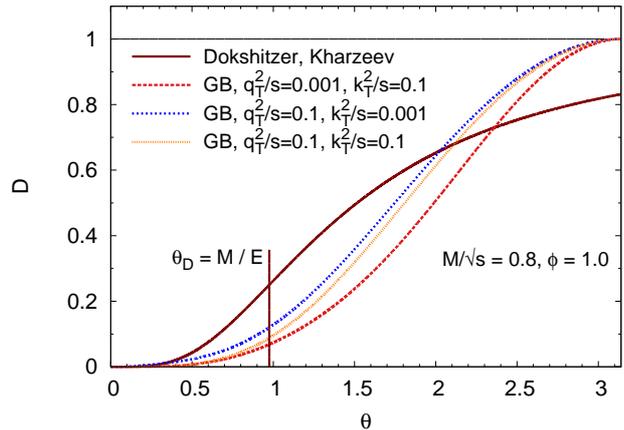}
\caption[Angle dependence of dead cone factor for $M/\sqrt{s} = 0.8$]{As Fig.~\ref{fig:dead_cone_gb_m01} but for $M/\sqrt{s} = 0.8$}
\label{fig:dead_cone_gb_m08}
\end{figure}

Figure~\ref{fig:dead_cone_gb_m01} shows the dead cone factor $\mathcal{D}_{\rm GB}$ as a function of $\theta$ for $M/\sqrt{s} = 0.1$ and $\phi = 1.0$  as well as different parameter sets for $k_\perp$ and $q_\perp$.
The curves of the different $q_\perp$-$k_\perp$ sets do not differ much and agree well with the suppression factor of Dokshitzer and Kharzeev. For large angles~$\theta$ all curves saturate at one, as it is expected. 
If $q_\perp$ is much smaller than $k_\perp$, the dead cone is more pronounced than that of Dokshitzer and Kharzeev. The dead cone factor is also strongly dependent on the the angle $\phi$ and the mass $M$. For small $\phi$ the suppression is considerably larger than $\mathcal{D}_{\rm DK}$, whereas for $\phi = \pi/2$ the suppression factor $\mathcal{D}_{\rm GB}$ does not depend on $k_\perp$ and $q_\perp$ anymore (both not plotted). For large masses $M$ the suppression of $\mathcal{D}_{\rm GB}$ is also significantly larger than $\mathcal{D}_{\rm DK}$ for the same mass, as shown in Fig.~\ref{fig:dead_cone_gb_m08} for $M/\sqrt{s} = 0.8$. 
This difference comes due to the approximation $E \gg M$ employed for $\mathcal{D}_{\rm DK}$. Since $\mathcal{D}_{\rm DK}$ is strictly only valid for small angles, the suppression factor does not go to one for large angles in contrast to our calculation, which is valid for all angles.

\begin{figure}
\centering
\includegraphics[width=\linewidth]{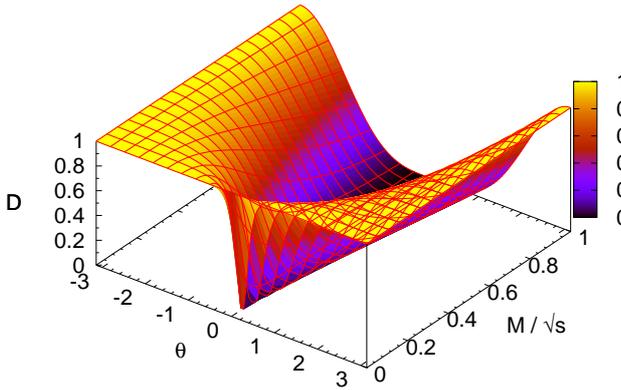}
\caption[Angle and mass dependence of dead cone factor]{The suppression factor $\mathcal{D}_{\rm AGMMU}$ from Eq.~\eqref{dead_cone_wo_kt_qt} as a function of $\theta$ and $M/\sqrt{s}$.}
\label{fig:dead_cone_3D_plot}
\end{figure}

To study the suppression factor $\mathcal{D}_{\rm GB}$ in more detail we employ the approximation $q_\perp \gg k_\perp$, but do not put any constraints on the heavy quark mass or the angle of the emitted gluon as has been done by Dokshitzer and Kharzeev. The suppression factor then reads
\begin{align}
\mathcal{D}_{\rm GB} &= \frac{P^g_M}{P^g_0} \simeq \frac{ \l[ \frac{ {\bf k}_\perp}{k_\perp^2+x^2M^2}  \r]^2 }{ \l[ \frac{ {\bf k}_\perp}{k_\perp^2}  \r]^2 }
= \left(1+\frac{x^2 M^2}{k_\perp^2}\right)^{-2} \breakeq
&= \left(1+\frac{M^2}{s\tan^{2}(\frac{\theta}{2})}\right)^{-2} =: \mathcal{D}_{\rm AGMMU}
\label{dead_cone_wo_kt_qt}
\end{align}
and does not depend on $k_\perp$ or $q_\perp$ anymore.
As a note, this is also the result that one would directly obtain from the result of Ref.~\cite{Abir:2011jb}, where the matrix element of the process $qQ\rightarrow qQg$ is calculated in Feynman gauge with slightly different approximations compared to the previous section. Therefore, we name this suppression factor in the very soft gluon approximation  in the following $\mathcal{D}_{\rm AGMMU}$ after the authors of Ref.~\cite{Abir:2011jb}.
As discussed in Ref.~\cite{Abir:2011jb}, it is valid in the full range of $\theta$---or rapidity of the emitted gluon---(i.e., $-\pi<\theta<+\pi$) and in the full range of $m=M/{\sqrt s}$ (i.e., $0<m<1$).
As can be seen in Fig.~\ref{fig:dead_cone_3D_plot} the suppression is rather narrow in $\theta$ for small $m$, but becomes very wide for large masses.
The valley around $\theta \simeq 0$ clearly indicates the presence of the dead cone in the forward direction with respect to the propagating heavy quark. In the backward region, $\theta \simeq \pm \pi$, the suppression factor saturates to unity. This suggests that the quark mass plays only a role in the forward direction where the energy of the quark becomes of the order of its mass.

As shown in Ref.~\cite{Abir:2011jb}, in the limit $M \ll \sqrt{s}$ and $\theta \simeq 0$, it is $\sqrt{s} \simeq 2E$ and $\tan(\theta/2)\simeq \theta/2$ and Eq.~\eqref{dead_cone_wo_kt_qt} reduces to the result of Dokshitzer and Kharzeev from Eq.~\eqref{dead_cone_factor}.


\subsection{Comparison of Gunion-Bertsch and exact cross sections}
\label{sec:comparison_gb_exact}
In this section we compare the improved GB matrix elements for the heavy quark process $qQ \arr qQg$ from Eq.~\eqref{gb_qQg_matrix_element_improved} to the exact matrix element calculated without any approximations. The exact matrix element for $qQ \arr qQg$ can be obtained from the process $q\bar{q} \arr Q\bar{Q}g$, which has been calculated in \Ref~\cite{Kunszt:1979iy}, by crossing the two anti-quarks. We checked numerically that it agrees with the exact matrix element for the light quark process $q q' \arr q q'  g$ \cite{Berends:1981rb}  if the mass of the heavy quark is set to zero.

The following comparison is done analogously to Ref.~\cite{Fochler:2013epa}, where the improved GB and exact matrix element for light partons have been compared. All calculations take into account the full kinematics for heavy quarks while ensuring energy and momentum conservation. 

Both the exact and GB matrix elements are divergent for infrared and collinear configurations. Usually these divergences are screened with a screening mass of the order of the Debye mass. However, since it is not straightforward to identify the propagators in the complex expression of the exact matrix element, we cut instead in the integration if the scalar product of any incoming or outgoing four-momenta is smaller than a cut-off $\Lambda^2$, $p_i \cdot p_j < \Lambda^2$ with $i,j = 1...5$, where $\Lambda^2$ is chosen to be proportional to the Debye mass, 
\begin{equation}
\label{eq:lambda_epsilon}
\Lambda^2 = \epsilon \, m_D^2 \ .
\end{equation}

The calculations in this section are done for a temperature of $T=400 \, {\rm MeV}$. The coupling is set constant, $\alpha_s = 0.3$, and the squared \cm energy to $s-M^2=18 T^2 \simeq 2.88 \, {\rm GeV}^{2}$, which is for massless particles the average thermal value. We determine $\Lambda$ from the usual gluon Debye mass for Boltzmann statistics as in Ref.~\cite{Fochler:2013epa}. For $n_f = 3$ at the given temperature and coupling, the Debye mass is $m_{D}^2 =8 \alpha_s (3+n_f) \, T^2 / \pi \simeq 0.73 \, {\rm GeV}^2$.

\begin{figure}
\centering
\includegraphics[width=\linewidth]{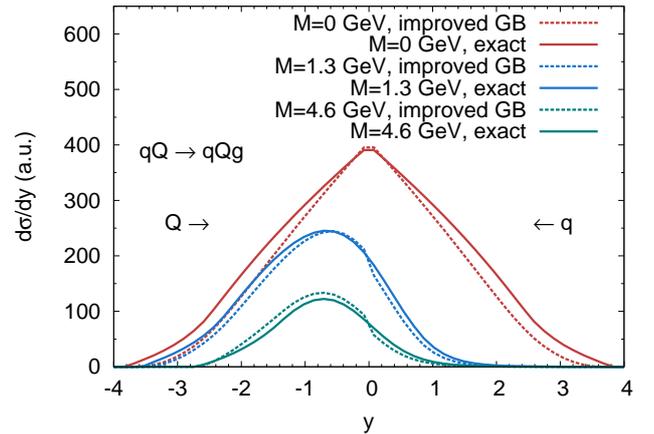}
\caption[Differential cross section in $y$ for $qQ \rightarrow qQg$]{Differential cross section ${\rm d} \sigma / {\rm d} y$ for heavy and light quark scattering, $qQ \rightarrow qQg$, calculated with the exact~\cite{Kunszt:1979iy} and improved GB~\eqref{gb_qQg_matrix_element_improved} matrix element for different heavy quark masses $M$.The calculations are done for $s-M^2 = 18 T^2$ and  $\epsilon = 0.001$. The incoming heavy quark $Q$ goes in the positive rapidity direction and the light quark $q$ in the negative direction.}
\label{fig:ds_dy_qQ}
\end{figure}

Figure~\ref{fig:ds_dy_qQ} compares the rapidity spectrum of the emitted gluon for the scattering of a heavy quark with a light quark, $qQ \rightarrow qQg$.
For a heavy quark mass of $M = 0 \unit{GeV}$, the process is in fact a scattering of two light quarks and, thus, symmetric. 
Furthermore, the gluon emission spectra for massive heavy quark scatterings are depicted for the charm and bottom masses. At forward rapidities---or small emission angles---the gluon emission spectrum is suppressed, which is a consequence of the dead cone effect (cf.~Sec.~\ref{sec:dead_cone_effect}).
Independent of the heavy quark mass, the agreement between the improved GB and exact matrix element is remarkably good.

\begin{figure}
\centering
 \includegraphics[width=\linewidth]{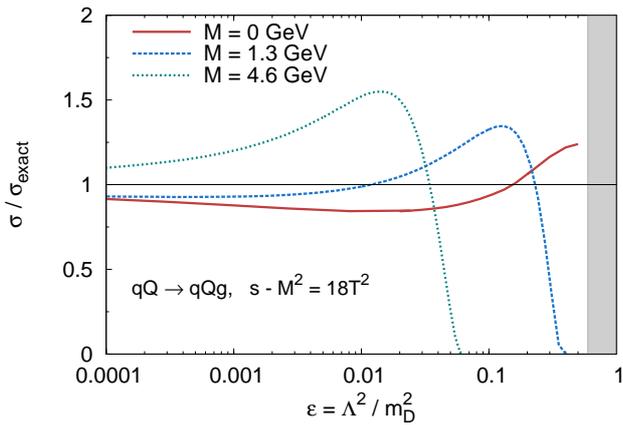}
\caption{Ratio of the total cross section $ \sigma $ of the improved GB matrix element to the exact result  for $qQ\rightarrow qQg$ as a function of the screening cut-off $\epsilon$ from Eq.~\eqref{eq:lambda_epsilon} for different heavy quark masses $M$.}
\label{fig:sigma_s_epsilon_22t_qQ}
\end{figure}

To reduce the screening effect as much as possible we set $\epsilon = 0.001$ in Fig.~\ref{fig:ds_dy_qQ}.
In Fig.~\ref{fig:sigma_s_epsilon_22t_qQ} we explore the dependence of the ratio of the total cross sections of the improved and exact matrix element on the cut-off parameter $\epsilon$.
The deviation from one for all $\epsilon$ is rather small for both light and heavy quarks. Note that the threshold, above which the whole phase space is cut away, is smaller for heavy quarks. Thus, these curves go back to zero for smaller $\epsilon$ compared to the light quark case where the gray box depicts the region with zero cross section.

We conclude that the improved GB approximation is a good approximation for studying radiative gluon bremsstrahlung processes for both light and heavy partons.


\subsection{Landau-Pomeranchuk-Migdal effect}
\label{sec:lpm}

The Landau-Pomeranchuk-Migdal (LPM) effect \cite{Landau:1953um,Migdal:1956tc} describes the suppression of medium-induced bremsstrahlung processes due to coherence effects of multiple scatterings.

For a jet traversing a medium radiation processes are induced by the medium.
If the wavelengths, or equivalently the formation times, of the radiated particles (gluons in QCD or photons in QED) are large compared to the time between subsequent interactions with the medium constituents, interference effects lead to a suppression of the radiation. In the QED case the time between two scattering processes is given by the mean free path of the jet. However, in QCD also the emitted gluon carries color charge and interacts, leading to a modification of its formation time and a more complicated suppression procedure \cite{Gyulassy:1993hr,Baier:1994bd,Baier:1996kr}.
 
The implementation of such a coherence effect is rather involved in a semi-classical transport model like BAMPS. To this end we implement an effective description of the LPM effect by allowing only processes that fulfill the relation \cite{Xu:2004mz, Fochler:2010wn, Uphoff:2014cba}
\begin{equation}
\label{lpm_constraint_x}
\lambda > X_\text{LPM} \,\tau \ ,
\end{equation}
where $\lambda$ is the mean free path of the considered particle and $\tau$ the  formation time of the emitted gluon. $X_\text{LPM}$ is a parameter of this effective prescription. While $X_\text{LPM}=0$ corresponds to the case of no LPM suppression and therefore is infrared divergent, $X_\text{LPM}=1$ allows only processes where the emitted gluon is already formed before the next scattering takes place, thus, forbids all possible interfering processes. We expect that a more sophisticated implementation of the LPM effect corresponds effectively to an $X_\text{LPM}$ that is between these two extreme cases, $0<X_\text{LPM}<1$. Recently, more sophisticated implementations of the LPM effect in Monte Carlo simulations have been introduced \cite{Zapp:2008af, Zapp:2011ya, ColemanSmith:2011wd, ColemanSmith:2012vr} and it would be interesting to study the impact of those on the BAMPS simulations. We leave this as an involved future project.

The constraint from Eq.~\eqref{lpm_constraint_x} can be ensured by multiplying the matrix element~\eqref{me_gb_hq} with the step function
\begin{equation}
\Theta\left( \lambda - X_\text{LPM} \,\tau \right)
\end{equation}
before integrating to calculate the cross section.
This makes the cross section---and thus also the rate---dependent on the mean free path, which in turn influences the mean free path. Hence, the actual mean free path of the jet must be determined in an iterative procedure,
\begin{equation}
 \label{iterative_MFP}
 \lambda = \lim_{i\rightarrow \infty} \lambda_{i} = \lim_{i \rightarrow \infty} \frac{v}{R_{22}+R_{23}(\lambda_{i-1})+R_{32}(\lambda_{i-1})} \ ,
\end{equation}
where $v$ is the velocity of the considered particle in the rest frame of the medium and $R_n$ the rate for process $n$. The rates for $2 \leftrightarrow 3$ are influenced by the LPM effect and, therefore, depend on the mean free path itself. However, for heavy quarks we will neglect $3 \arr 2$ processes since their rates are much smaller than the other processes.

The formation time of a radiated gluon of a heavy quark with mass $M$ is given in the center of mass frame by \cite{uphoff_phd}
\begin{align}
\label{formation_time_gluon_radiation_lab_12}
 \tau_{\rm cms} \simeq \frac{\omega}{k_\perp^2 + x^2 M^2} \ .
\end{align}
It is important to consider the formation time and the mean free path in the same frame of reference. The boosting from frame to frame is done analogously as outlined in Ref.~\cite{Fochler:2010wn}. There, the formation time of gluon emission off massless partons is calculated in the frame in which the emitted gluon is purely transverse. According to the uncertainty principle the formation time reads $\tau_{\rm trans} = 1/k_\perp$. This result can be generalized to heavy quarks by boosting Eq.~\eqref{formation_time_gluon_radiation_lab_12} to this frame,
\begin{align}
\label{formation_time_gluon_radiation_trans_23}
 \tau_\text{trans} = \frac{k_\perp}{k_\perp^2 + x^2 M^2} \ .
\end{align}
As a note, the whole treatment of the LPM cut-off for heavy quarks is in line with the implementation for light partons \cite{Xu:2004mz, Fochler:2010wn, Uphoff:2014cba} and reduces to the latter if the heavy quark mass $M$ is set to zero.

\subsection{Phase space constraints due to kinematics and the Landau-Pomeranchuk-Migdal effect}
\label{sec:phase_space_lpm}

The allowed phase space of the emitted gluon is constrained by the LPM cut-off and by kinematics. The constraint from kinematics due to energy and momentum conservation demands
\begin{align}
\label{23phase_space_kin}
 k_{\perp}  < \frac{\sqrt{s} (1 - m^2)}{2 \cosh y } \ ,
\end{align}
where we defined $m = M / \sqrt{s}$ as the heavy quark mass $M$ scaled by the \cm energy~$\sqrt{s}$. 

The implementation of the LPM cut-off allows only interactions if $\lambda > X_\text{LPM} \,\tau = X_\text{LPM} \,\gamma \tau_\text{trans}$, where $\gamma$ denotes the boost from the lab frame to the frame in which the emitted gluon in purely transverse. By employing Eq.~\eqref{long_momentum_frac_x_org} and taking into account the boost between the different reference frames, in which $\lambda$ and $\tau_\text{trans}$ are calculated, this constraint can be rewritten in terms of $y$ and $k_\perp$, \cite{uphoff_phd}
\begin{align}
\label{23phase_space_lpm}
  k_\perp > X_\text{LPM}\frac{\cosh y + A \sinh y}{B \l( 1 + m^2 e^{2y} \r)}
\end{align}
with $A = \beta_{\rm cms}\cos \theta$ and $B = \lambda \sqrt{1-\beta^{2}_{\rm cms}}$, where $\beta_{\rm cms}$ denotes the boost from the lab to the center-of-mass frame and $\theta$ the angle between the the boost from lab to center-of-mass frame and the boost from center-of-mass frame to the frame in which the emitted gluon is purely transverse \cite{uphoff_phd}.

This relation only holds if the heavy quark is the incoming particle number 1 that flies in the positive $z$ direction. If the heavy quark is particle 2, all the findings of this section must be substituted with $y \arr -y$ due to the symmetry of the process.

This implies that the available phase space of the emitted gluon  in the $y$-$k_\perp$-plane in the center-of-mass frame is enveloped between the two curves
\begin{align}
  f_\text{kin}(y)  &= \frac{\sqrt{s} (1 - m^2)}{2 \cosh y } \label{eq:chap_BAMPS:kt_of_y_LPM_constraint} \qquad\qquad
\nonumber \\
  f_\text{lpm}(y) &= X_\text{LPM} \frac{\cosh y + A \sinh y}{B \l( 1 + m^2 e^{2y} \r)} \ ,
\end{align}
which stem from Eqs.~\eqref{23phase_space_kin} and \eqref{23phase_space_lpm}, respectively. 

\begin{figure}
  \begin{center}
    \includegraphics[width=\linewidth]{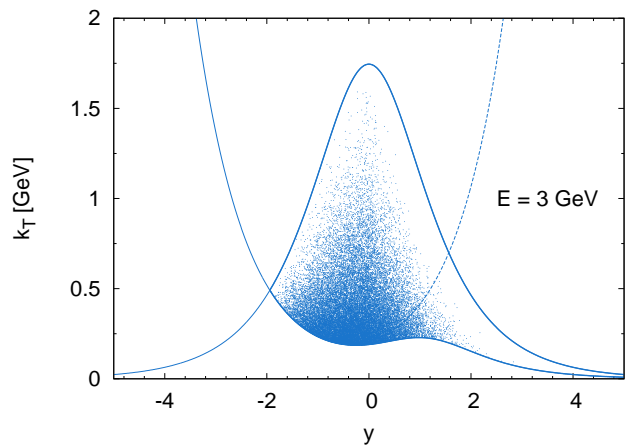}
    \caption[Phase space sampling of emitted gluon off heavy quark with $E=3 \unit{GeV}$.]{Phase space sampling in the $y$-$k_\perp$-plane of the emitted gluon in the center-of-mass frame for a heavy quark with $E=3 \unit{GeV}$ and mass $M=1.3 \unit{GeV}$.  The heavy quark comes from the left and has a momentum of $p_{1} = (E, 0, 0, \sqrt{E^2-M^2})$. The medium particle is taken as $p_{2} = (3 T, 0, 0, -3 T)$ with $T = 0.4 \unit{GeV}$.
The upper envelope curve is the constraint due to kinematics, $f_\text{kin}(y)$, while the lower line is the LPM constraint,  $f_\text{lpm}(y)$. Allowed phase space is the area between these two curves. For comparison, the dashed line shows the LPM constraint for a massless quark. Each of the $100\,000$ dots represents one sampled phase space point from BAMPS. The mean free path is taken to be $\lambda = 1 \unit{fm}$ and  $X_\text{LPM}=1$.
}
    \label{fig:kt_y_phase_space_sampling} 
  \end{center}
\end{figure}

In Fig.~\ref{fig:kt_y_phase_space_sampling} the available phase space in the $y$-$k_\perp$-plane is illustrated. 
The drawn curves show the envelope functions $f_\text{kin}(y)$ and $f_\text{lpm}(y)$. The dashed line represents for comparison the LPM constraint for a massless quark. In this case the LPM effect serves as a lower cut-off in $k_\perp$. In contrast, the emission from a massive quark with very small $k_\perp$ is allowed at very forward rapidity because of the mass dependence of the formation time, although the cross section in this region is very small due to the $(1- \bar x)$ factor and the $x^2 M^2$ term in the denominator of the matrix element from Eq.~\eqref{gb_pgm_radiation_spectrum} (dead cone effect).
This interplay between the LPM and dead cone effect prevents a divergence of the cross section for small $k_\perp$ in the case of heavy quark scatterings.

Each dot depicts a sampled phase space point from BAMPS for the given parameters. It can be nicely seen that all sampled phase space points obey the phase space constraints from Eqs.~\eqref{23phase_space_kin} and \eqref{23phase_space_lpm}. The density of the points is a measure for the value of the matrix element in this phase space region. The density increases with smaller $k_\perp$ and is largest around mid-rapidity. At positive rapidities the density is lower than at negative rapidities, which is due to the dead cone effect (see Sec.~\ref{sec:dead_cone_effect}).


\section{Elastic vs.\ radiative energy loss in a static medium}
\label{sec:eloss_radiative}

Before turning to full heavy-ion collisions we study in this section the energy loss of light and heavy quarks in a static thermal medium with a temperature of $T = 400 \unit{MeV}$. In the following we set the LPM parameter to $X_\text{LPM}=1$, but explore at the end of this section the energy loss dependence on this parameter.

\begin{figure}
\centering
\includegraphics[width=\linewidth]{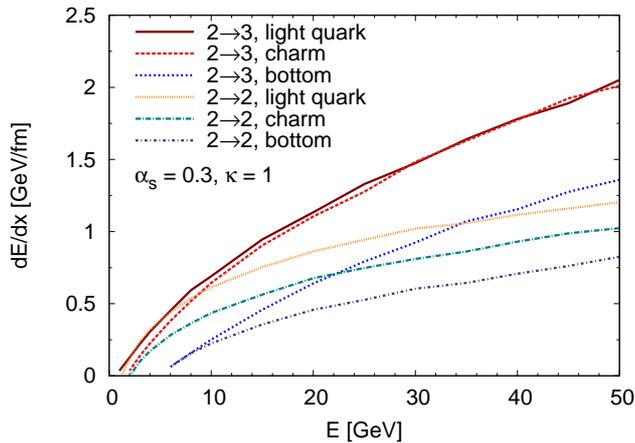}
\caption[Elastic and radiative energy loss of light and heavy quarks with constant coupling and $\kappa_t = 1$]{Elastic and radiative energy loss per unit length of a light quark ($M=0\unit{GeV}$), a charm quark ($M=1.3\unit{GeV}$), and a bottom quark ($M=4.6\unit{GeV}$) traversing a static thermal medium with temperature $T=0.4\unit{GeV}$. The curves are calculated with constant coupling and Debye mass prefactor $\kappa_t = 1$. The mean free path that enters in the LPM cut-off is determined iteratively.}
\label{fig:dedx_mass_asConst}
\end{figure}

Figure~\ref{fig:dedx_mass_asConst} depicts the elastic and radiative energy loss of light, charm, and bottom quarks for a constant coupling ($\alpha_s=0.3$) and standard Debye screening ($\kappa_t=1$). %
The radiative energy loss is larger than the elastic energy loss for all quark masses. While both have similar sizes at small energy, they differ by about a factor of two at larger energies for all flavors.
The mass hierarchy is visible for both the elastic and radiative energy loss---heavy quarks lose less energy. However, between light and charm quarks the difference for the radiative energy loss is only marginal. As we have mentioned in Sec.~\ref{sec:lpm}, the LPM cut-off
suppresses the gluon emission at small angles and partly overshadows the dead cone of heavy quarks. The charm quark dead cone 
is rather small at larger energies. Consequently, the energy loss is mostly influenced by the LPM suppression, which renders the charm and light quark curves to be rather similar. 

\begin{figure}
\centering
\includegraphics[width=1.0\linewidth]{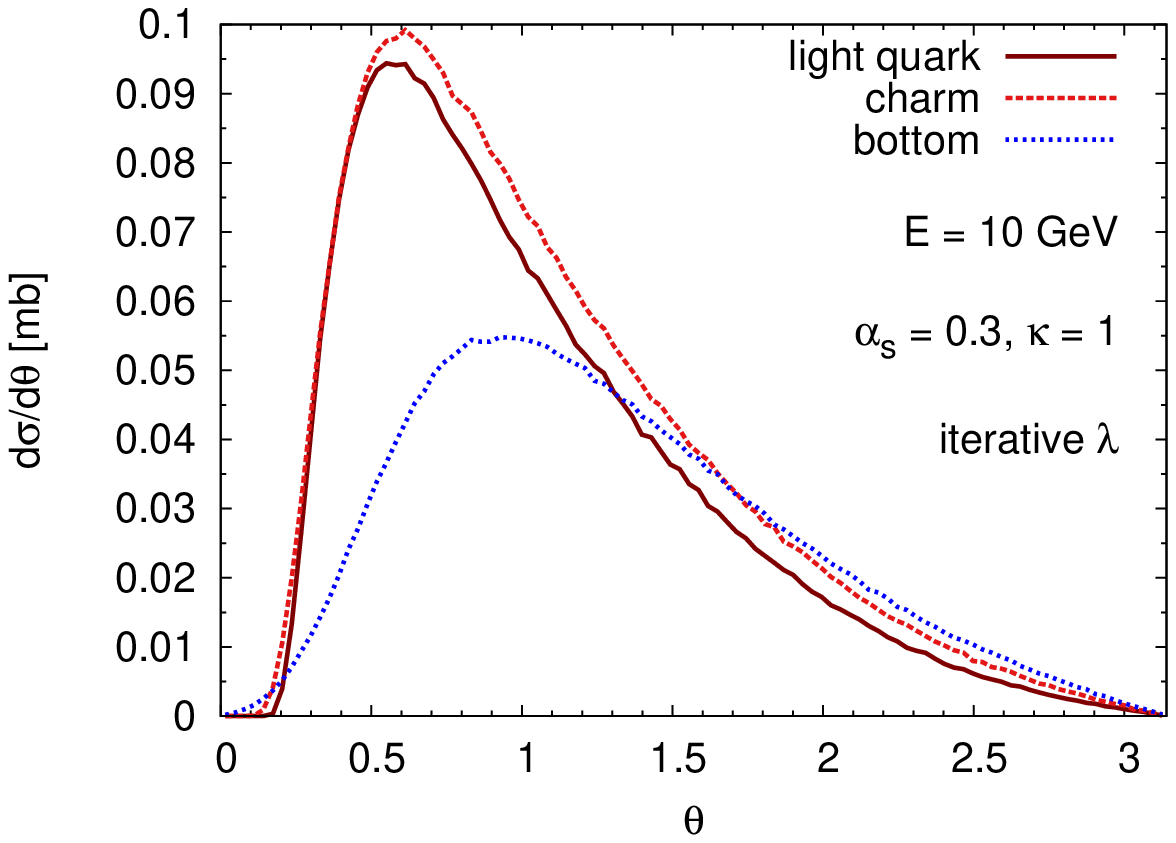}

\includegraphics[width=1.0\linewidth]{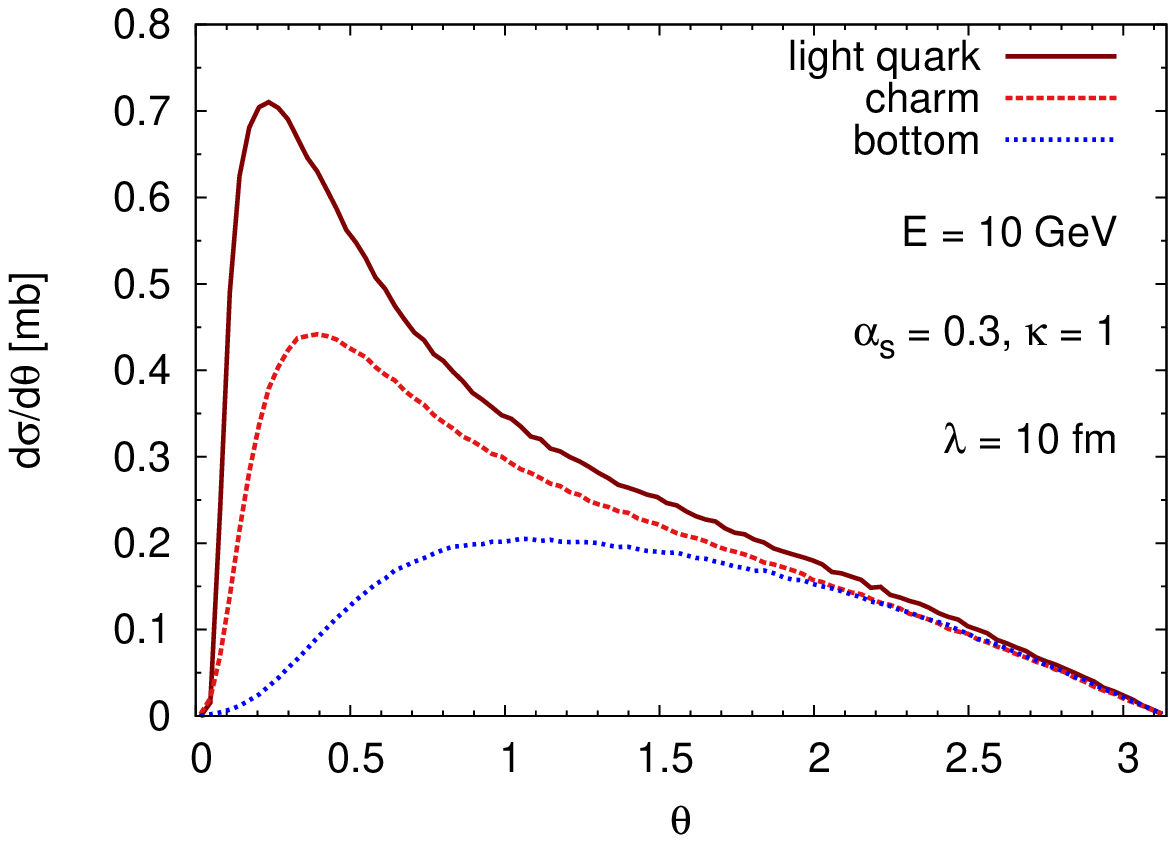}
\caption[Angle distribution of emitted gluon for light, charm, and bottom jets]
{Average differential cross section $\d \sigma / \d \theta$ of a light, charm, and bottom quark jet with energy $E=10\unit{GeV}$ as a function of the angle $\theta$, at which the gluon is emitted in the lab frame with respect to the jet direction. For the upper plot the mean free path that enters in the LPM cut-off is determined iteratively, while for the lower plot it is set by hand to the large value of $\lambda = 10 \unit{fm}$.}
\label{fig:dsigma_dtheta_23}
\end{figure}

This reasoning can be nicely verified by looking at the angular distribution of the gluon emitted off a light and charm quark jet. In the upper panel of Fig.~\ref{fig:dsigma_dtheta_23} the differential cross section is depicted as a function of the gluon emission angle with respect to the jet for light, charm, and bottom quarks. 
The distributions of light and charm quarks are very similar. Small angles are suppressed due to the LPM cut-off, but no additional dead cone suppression for charm jets can be seen.

For illustration, in the lower panel of Fig.~\ref{fig:dsigma_dtheta_23} the mean free path that enters in the LPM cut-off is not determined iteratively (which is the default implementation in BAMPS, see Sec.~ref{sec:lpm} and Ref.~\cite{Fochler:2010wn}), but set by hand to a large value of $\lambda = 10 \unit{fm}$. Hence, the LPM suppression is reduced and smaller angles are allowed. Consequently, the suppression due to the dead cone effect becomes visible again, which results in a stronger suppression of charm quarks at small angles compared to light quarks. 

In contrast to charm quarks, the dead cone for bottom quarks is more pronounced and is even visible for an iteratively calculated mean free path. Hence, also the energy loss in Fig.~\ref{fig:dedx_mass_asConst}  is considerably smaller than for lighter quark flavors. The similar energy loss of light and charm quarks could be an explanation why the measured nuclear modification factors of charged hadrons and $D$ mesons  in heavy-ion collision have the same values. We discuss this in more detail in Sec.~\ref{sec:results_open_hf_radiative}.

Similar features can be seen for a running coupling (not shown). Only the energy dependence of elastic and radiative energy loss changes slightly since the running coupling decreases with increasing momentum transfer.

\begin{figure}
\centering
\includegraphics[width=\linewidth]{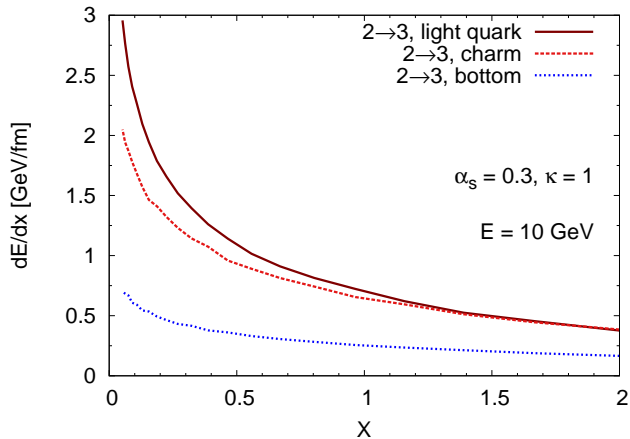}
\caption[Influence of LPM effect on radiative energy loss of light and heavy quarks]{Dependence of the radiative energy loss of light, charm, and bottom quarks with $E=10\unit{GeV}$ on $X_\text{LPM}$, which denotes a prefactor in the LPM cut-off, $\Theta\left( \lambda - X_\text{LPM} \,\tau \right)$.}
\label{fig:dedx_mass_asConst_Xlpm}
\end{figure}

The previous calculations have been done for the LPM parameter $X_\text{LPM}=1$.
Figure~\ref{fig:dedx_mass_asConst_Xlpm} depicts the $X_\text{LPM}$ dependence of the energy loss for light, charm, and bottom quarks with an energy of $E=10\unit{GeV}$.
The total cross section depends logarithmically on $X_\text{LPM}$. Since the energy loss per collision is only mildly influenced by $X_\text{LPM}$, the energy loss per unit length is also logarithmically dependent on $X_\text{LPM}$, $\d E/\d x \sim \ln 1/X_\text{LPM} $. As discussed above, the energy loss of light quarks and charm quarks is comparable for $X_\text{LPM}=1$ (cf.\ Fig.~\ref{fig:dedx_mass_asConst}) because an LPM dead cone overshadows the mass dead cone of the charm quark. For $X_\text{LPM}<1$ the impact of the LPM cut-off is reduced and the energy loss of charm quarks becomes smaller than that of light quarks, which is in accordance with the expectations from the dead cone effect.


\section{Elliptic flow and nuclear modification factor from elastic and radiative processes}
\label{sec:results_open_hf_radiative}

In Ref.~\cite{Uphoff:2011ad,Uphoff:2012gb} we saw with BAMPS that binary collisions alone cannot explain the experimental heavy flavor data, neither at RHIC nor at LHC. However, if the binary cross sections are scaled with a phenomenological $K$ factor, both the elliptic flow \vt and nuclear modification factor \raa of heavy flavor particles can be described at RHIC and LHC. We concluded that missing contributions such as radiative processes or quantum effects play an important role.

In this section we explicitly add the radiative contributions and study if binary and radiative processes together can explain the experimental data. In Sec.~\ref{sec:results_light_parton} the focus is put on the similarities and differences of light and heavy flavor particles. Furthermore, the influence of the dead cone effect is investigated. To treat light and heavy flavor partons on the same footing, we use for all species the standard Debye screening. In Sec.~\ref{sec:23_hf_kappa} we explicitly study the effect of the improved screening procedure and confront various heavy flavor calculations with experimental data.

Heavy quarks and high-energy light partons studied in this section are set on top of fully simulated BAMPS background events to enhance the statistics of these rare probes. This treatment is in line with full BAMPS simulations, but neglects medium response effects. In this section we use solely background events calculated with the original Gunion-Bertsch cross section \cite{Gunion:1981qs}.  They correctly reproduce bulk medium properties such as the measured elliptic flow of charged hadrons \cite{Xu:2007jv,Fochler:2011en}. We checked that background events calculated with the improved Gunion-Bertsch cross section, do not change the \raa results presented in this section due to a similar number of scattering centers. However, the light parton elliptic flow with the improved Gunion-Bertsch cross section is slightly reduced, which in turn would also slightly decrease the elliptic flow of heavy quarks. In summary, the results in this section are calculated with a bulk medium that correctly features the experimentally measured properties and, thus, is appropriate to study the suppression and flow of hard probes. Furthermore, it is the same medium evolution in which previous heavy flavor BAMPS calculations \cite{Uphoff:2011ad,Uphoff:2012gb} have been performed.

\subsection{Comparison between heavy and light flavor}
\label{sec:results_light_parton}

Since the improved screening procedure with setting the screening mass prefactor to $\kappa_t = 0.2$ has only been derived for massive quarks and we want to treat heavy and light flavor on the same footing in order to make proper comparisons, we employ in this section the standard Debye screening for both light and heavy partons. Nevertheless, in the next section we investigate the influence of the improved screening procedure on heavy quark observables. It would be also interesting to extend the derivation of the improved screening procedure from comparisons to HTL calculations to the light parton sector and determine the value of $\kappa_t$ for light particles.

It is important to note that the same cross sections are employed for light and heavy partons. That is, if the mass in the heavy flavor cross sections for elastic or radiative processes is set to zero the cross sections of light quarks \cite{Uphoff:2014cba} are obtained. Also the LPM cut-off and the treatment of all the kinematics are implemented in BAMPS consistently for massive and massless particles.
Since $3\arr 2$ processes have only minor relevance for the energy loss of high-energy particles, we neglect those processes in the following study.

The fragmentation of heavy quarks to $D$ and $B$ mesons is performed with Peterson fragmentation \cite{Peterson:1982ak} as for previous studies. 
Light partons are fragmented to charged hadrons by using the AKK fragmentation functions \cite{Albino:2008fy}, which are obtained from global fits to data.

As discussed in Sec.~\ref{sec:lpm}, the way how the LPM effect is implemented in BAMPS forbids too many of the radiative processes, leading to a destructive interference that is too strong. A more sophisticated treatment of the LPM effect should allow more processes. We introduced a first step in this direction in Sec.~\ref{sec:lpm} by lowering the formation time of the emitted gluon by a factor $X_\text{LPM}$, which effectively allows more radiative processes for $0 < X_\text{LPM}<1$.

\begin{figure}
	\centering
\includegraphics[width=\linewidth]{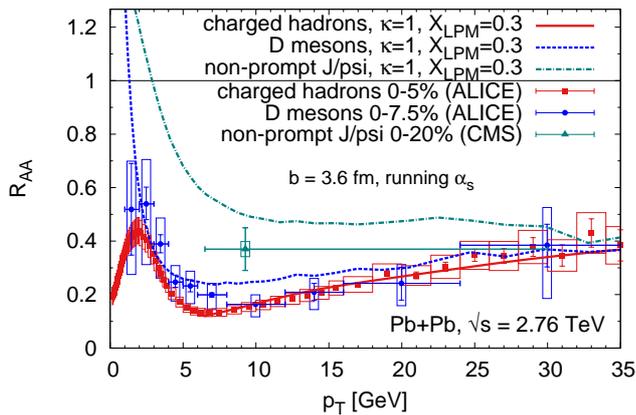}
	\caption[Charged hadron, $D$ meson, and non-prompt \jpsi $R_{AA}$ at LHC]{Nuclear modification factor $R_{AA}$ of charged hadrons, $D$ mesons, and non-prompt \jpsi at LHC for $X_\text{LPM}=0.3$  in comparison to data \cite{Abelev:2012hxa,Grelli:2012yv}. Note that the impact parameter $b=3.6\unit{fm}$ represents approximately a centrality class of 0-10\,\%, while the data is only available for slightly different centrality classes.}
	\label{fig:raa_data_npjpsi_dmeson_charged_hadron_X03}
\end{figure}

In Ref.~\cite{Uphoff:2014cba} we found that the best agreement with the neutral pion nuclear modification factor data at RHIC is given for  $X_\text{LPM}=0.3$. Simultaneously, the $R_{AA}$ of charged hadrons at LHC is nicely reproduced, not only the magnitude of suppression but also the shape and especially the rise with $p_T$ (see also Fig.~\ref{fig:raa_data_npjpsi_dmeson_charged_hadron_X03}). 
For the same value also a significant light parton elliptic flow is observed due to a small shear viscosity over entropy density ratio of about $0.2-0.3$ at late stages. However, it is important to mention that the exact value of $X_\text{LPM}$ is not theoretically motivated and, thus, a free parameter. Nevertheless, we expect that a more sophisticated LPM implementation boils effectively down to an $X_\text{LPM}<1$. It is planned for the future to improve the implementation of the LPM effect in BAMPS and study the impact on observables such as the \raa.

If we set the same value $X_\text{LPM}=0.3$ also for heavy quarks, the experimental data of the $D$ meson nuclear modification factor is also nicely reproduced, as can be seen in Fig.~\ref{fig:raa_data_npjpsi_dmeson_charged_hadron_X03}. Since we employ the standard Debye screening in this section to heavy quark interactions as well, both heavy and light partons are treated consistently. Although the dead cone effect is present in our matrix element, the \raa of $D$ mesons and charged hadrons are comparable, as it is also the case for the data. There are two reasons for that: first, as described in detail in Sec.~\ref{sec:eloss_radiative}, the LPM cut-off produces a second dead cone that overlays the dead cone due to the heavy quark mass and effectively annihilates its influence. This leads to a similar suppression of light and heavy quarks. However, gluons are still stronger suppressed. The second reason lies in the fragmentation process. The $D$ meson \raa is not really modified by the fragmentation, whereas the fragmentation of light quarks and gluons shifts the light hadron \raa closer to the light quark curve (see Ref.~\cite{Uphoff:2014cba} for details) and, thus, also to the $D$ meson curve. Consequently, due to the interplay between these two effects, both the $D$ meson and charged hadron \raa take very similar values. 

Recently, the same observation was made by Djordjevic \cite{Djordjevic:2013pba} within an extension of the WHDG framework. Also in this model, the suppression of light and charm quarks is comparable, whereas gluons are more suppressed. However, analogously to what we have observed, mass effects in the fragmentation function result in a similar suppression of charged hadrons and $D$ mesons.

In Fig.~\ref{fig:raa_data_npjpsi_dmeson_charged_hadron_X03} we depict besides the charged hadron and $D$ meson \raa for $X_\text{LPM}=0.3$ also the  \raa of non-prompt \jpsi for the same parameter.
Non-prompt \jpsi, which stem from $B$~mesons, are less suppressed than $D$ mesons or charged hadrons. Because of the large mass of bottom quarks, the dead cone due to the mass is larger than the dead cone due to the LPM cut-off. This leads to a smaller suppression of bottom quarks compared to charm quarks. The experimental data of non-prompt \jpsi is slightly smaller than our curve but still larger than the $D$ meson data (however, note the slightly different centrality classes).

\subsection{Open heavy flavor at RHIC and LHC with improved screening}
\label{sec:23_hf_kappa}

In the previous section we employed the standard Debye screening to ensure a similar treatment of light and heavy partons. In contrast, we study in this section the effect of the improved screening procedure ($\kappa_t = 0.2$) on heavy flavor observables.

\begin{figure}
	\centering
\includegraphics[width=\linewidth]{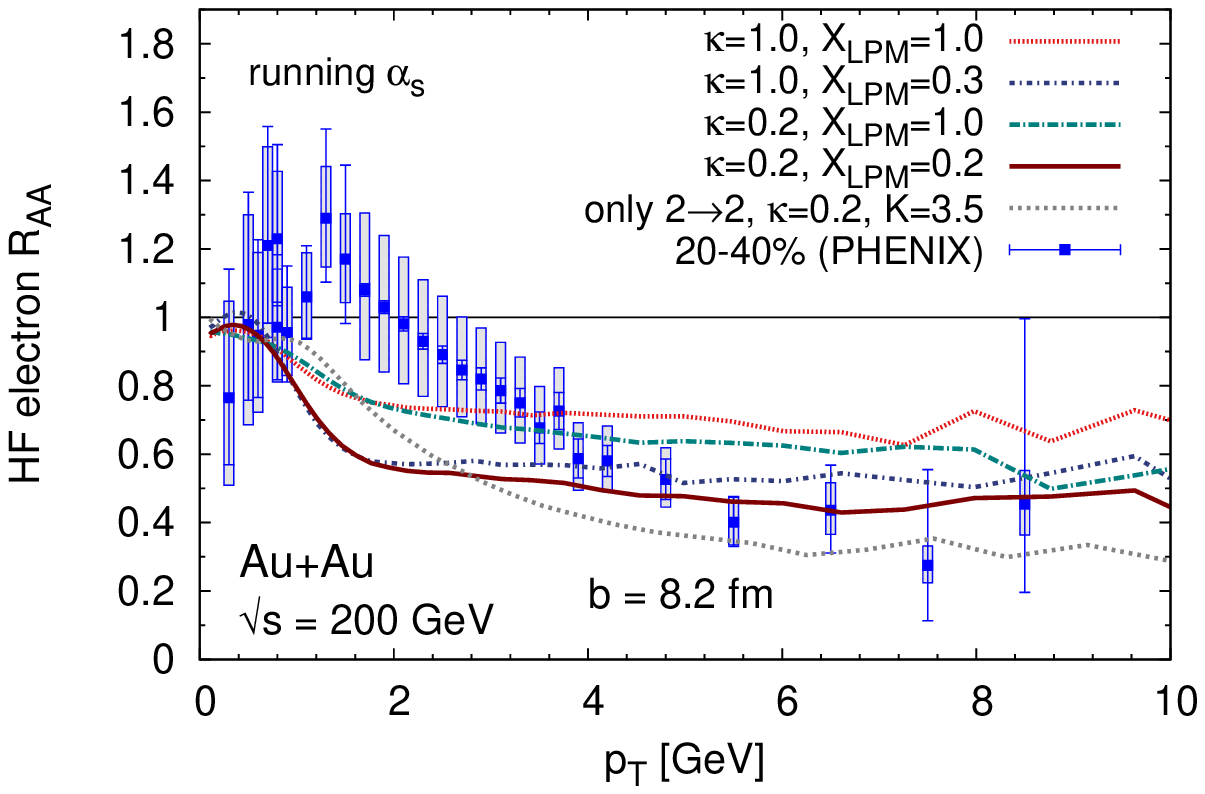}

\includegraphics[width=\linewidth]{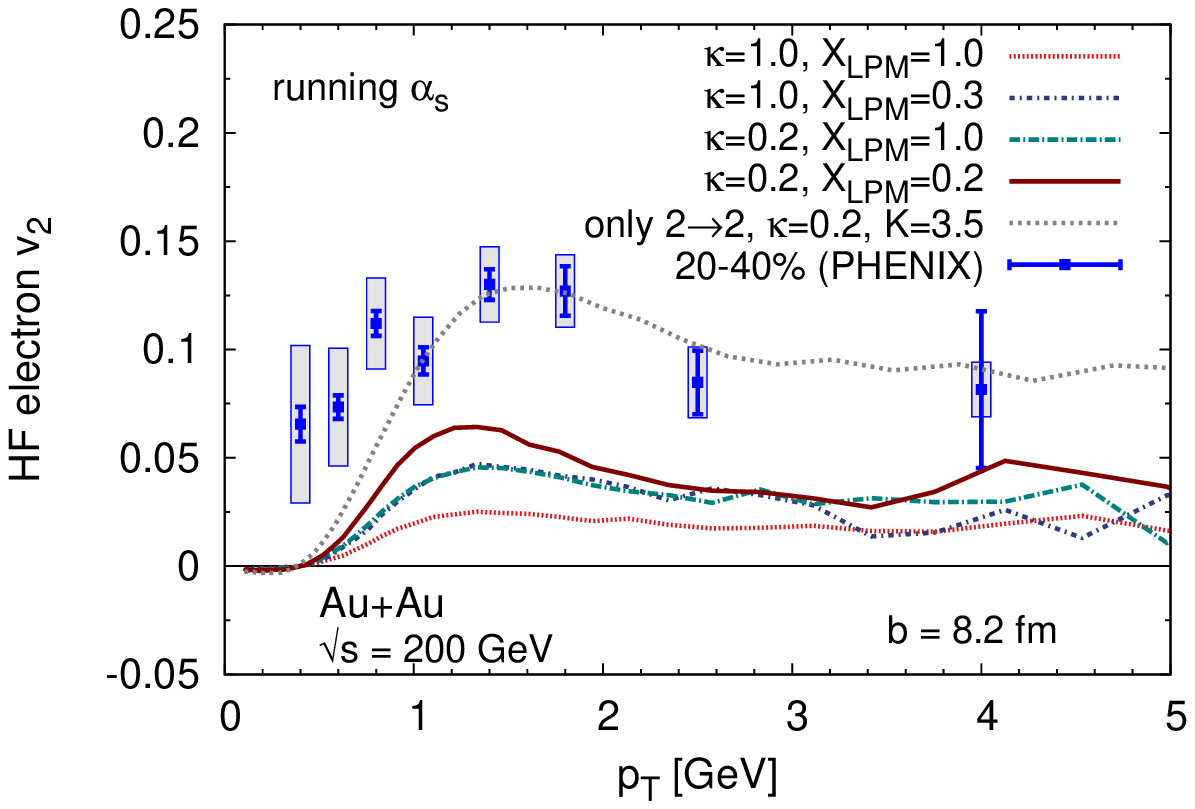}
	\caption[Heavy flavor electron $v_2$ and $R_{AA}$ at RHIC]{Nuclear modification factor $R_{AA}$ (top) and elliptic flow $v_2$ (bottom)  of heavy flavor electrons in non-central events at RHIC for running and different values of the screening prefactor $\kappa_t$ and the LPM parameter $X_\text{LPM}$. For comparison, data \cite{Adare:2010de} and the curve with only scaled ($K=3.5$) binary collisions \cite{Uphoff:2012gb} are shown.}
	\label{fig:raa_v2_23_data_electrons_rhic}
\end{figure}

Figure~\ref{fig:raa_v2_23_data_electrons_rhic} depicts the \raa and \vt of heavy flavor electrons in non-central RHIC collisions for various parameter sets.
The suppression with standard Debye screening ($\kappa_t=1$) and $X_\text{LPM}=1$ is smaller than the data, since too many radiative processes are forbidden due to the LPM cut-off. Lowering the LPM parameter to  $X_\text{LPM}=0.3$ as in the previous section gives a good agreement with the \raa data of heavy flavor electrons at large $p_T$, while the neutral pions at RHIC are also well described for the same parameter \cite{Uphoff:2014cba}. 

However, for both values of $X_\text{LPM}$ the elliptic flow of heavy flavor electrons is significantly smaller than the data. Qualitatively, this effect is also present for light quarks. Although the gluons build up a sizeable elliptic flow at RHIC, the light quarks have a significantly smaller flow \cite{Uphoff:2014cba} due to the smaller color factor ($4/9$ smaller compared to gluons) in the cross section. The elliptic flow of heavy quarks cannot be larger than that of light quarks for the same parameters. Hence, the \vt of electrons that is comparable to that of charged hadrons cannot be described.

Setting $\kappa_t = 0.2$ leads to a large binary cross section and a small mean free path of the heavy quark. Due to the implementation of the LPM effect (see Sec.~\ref{sec:lpm}), most of the radiative processes are forbidden. Thus, the total rate or energy loss for running coupling and improved screening is completely dominated by the binary processes. Since effectively no radiative processes are allowed, also the \raa is above the data for large~$p_T$.

In Ref.~\cite{Uphoff:2014cba} we found that the best description of the \raa data of light particles is given with $X_\text{LPM}=0.3$, for which also the \raa of $D$ mesons  with a standard Debye screening is well described (see previous section). If we employ the improved screening procedure, the best agreement with the \raa data at large $p_T$ is found for $X_\text{LPM}=0.2$, which is depicted in Fig.~\ref{fig:raa_v2_23_data_electrons_rhic}. It is worth noting that the value of $X_\text{LPM}$ that gives a good agreement with the improved screening lies in the same range as that for the standard Debye screening. However, we want to emphasize again that the exact value of $X_\text{LPM}$ is a free parameter, although a value smaller than one is expected from theoretical considerations. We expect that a more sophisticated treatment of the LPM effect makes such $X_\text{LPM}$ factors obsolete.

\begin{figure}
\centering
\includegraphics[width=1.0\linewidth]{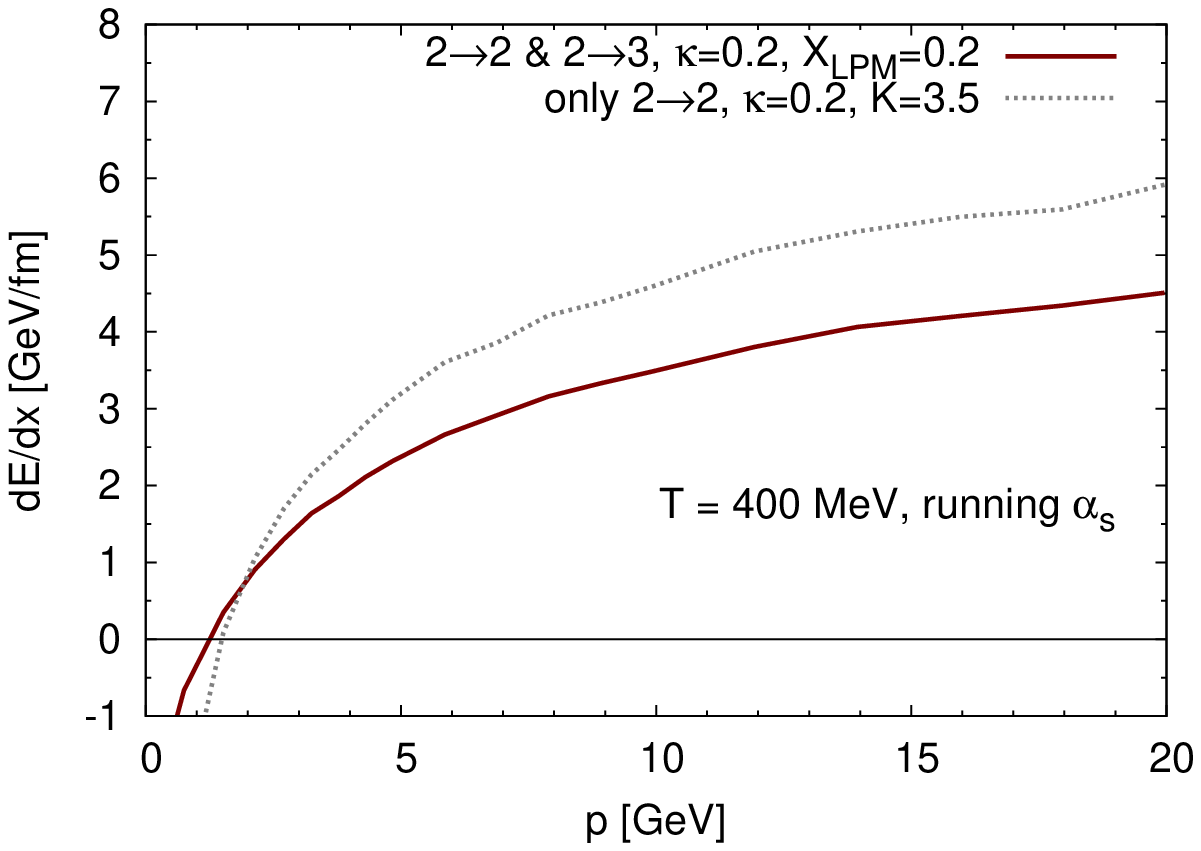}

\includegraphics[width=1.0\linewidth]{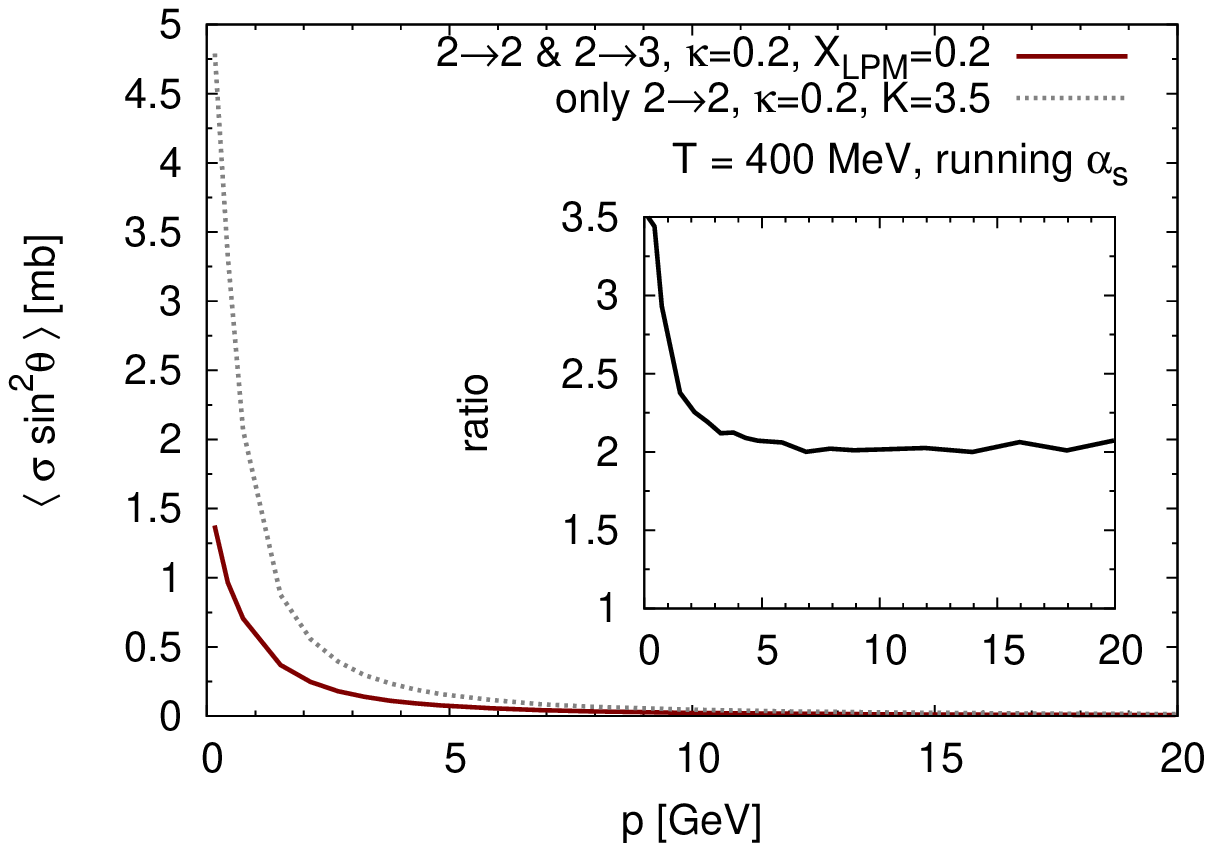}

	\caption[Energy loss and transport cross section of charm quarks in static medium]{Energy loss $\d E/\d x$ (upper panel) and transport cross section, which we define as $\left\langle \sigma \sin^2\theta\right\rangle$, where $\theta$ is the deflection of the charm quark in the lab frame, (lower panel) of charm quarks with momentum $p$ in a static medium with a temperature $T = 400 \unit{MeV}$. A running coupling and improved Debye screening are employed. The first curve includes binary and radiative collisions with $X_\text{LPM}=0.2$ and the second curve only binary collisions with $K=3.5$. 
  }
	\label{fig:transport_cs_23}
\end{figure}

For comparison, we also show the curve from Ref.~\cite{Uphoff:2012gb}, which was obtained by allowing only binary collisions and scaling the cross sections with $K=3.5$. It gives a slightly stronger suppression than the curve with $X_\text{LPM}=0.2$ at large~$p_T$ and a slightly weaker suppression at small~$p_T$. This reflects the observation made for the energy loss in a static medium (see also Fig.~\ref{fig:transport_cs_23}): including radiative processes is not the same as scaling the binary cross sections with a constant $K$~factor since both energy loss distributions have distinguished energy dependencies.

For the elliptic flow, the picture looks very different (see lower panel of Fig.~\ref{fig:raa_v2_23_data_electrons_rhic}). None of the curves that include radiative interactions can describe the elliptic flow data. In contrast, the scenario with the scaled binary interactions gives a sizeable \vt, although its \raa is comparable to the curve with $X_\text{LPM}=0.2$. This similarity in the \raa and the large deviation for the \vt, simultaneously, are at first sight surprising. However, the mechanisms responsible for the two observables are very different. For the \raa, the energy loss is the important quantity, while \vt is most effectively build up by a large transport cross section.

In Fig.~\ref{fig:transport_cs_23} we depict the energy loss and transport cross section, which we define as $\left\langle \sigma \sin^2\theta\right\rangle$, of charm quarks in a static medium for the two scenarios: only binary interactions scaled with $K=3.5$ and binary as well as radiative processes with $X_\text{LPM}=0.2$.
Although the energy loss of both scenarios is on the same order,\footnote{The scaled binary cross section has a slightly larger energy loss, which is reflected in a slightly smaller \raa at large transverse momentum in Fig.~\ref{fig:raa_v2_23_data_electrons_rhic}.} the transport cross section deviates strongly. This leads to a very similar \raa, but a large \vt for the case with only (scaled) binary interactions. However, a more detailed study (not shown here) reveals that the angular distribution is rather similar for elastic and radiative processes. Only the cross sections of both scenarios differ substantially. Due to the large $K=3.5$ factor, the scenario with only scaled elastic collisions has a much larger cross section than the scenario with elastic and radiative processes. Since the energy loss per collision for radiative processes is significantly larger than for elastic interactions, the overall energy loss per unit length is similar. Nevertheless, the large cross section for the scenario with only scaled binary collisions produces a large \vt that is in agreement with the data.

\begin{figure}
	\centering
\includegraphics[width=\linewidth]{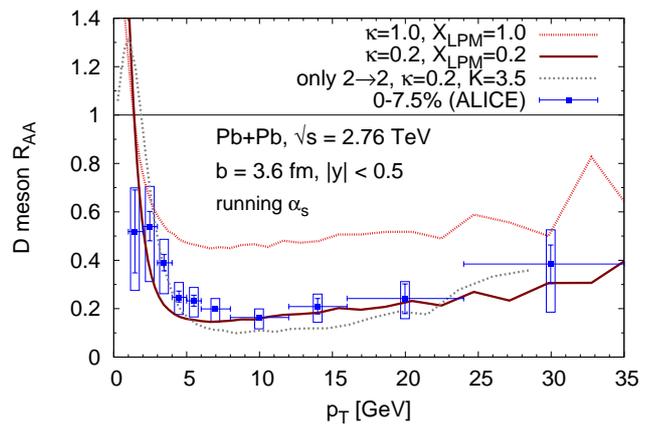}
	\caption[$D$ meson $R_{AA}$ at LHC]{Nuclear modification factor $R_{AA}$ of $D$ mesons at LHC for $b=3.6\unit{fm}$ with data \cite{Grelli:2012yv}. A running coupling is employed and either the parameter set $\kappa_t = 1$, $X_\text{LPM}=1$ or $\kappa_t = 0.2$, $X_\text{LPM}=0.2$. For comparison, the scaled binary collision curve from Ref.~\cite{Uphoff:2012gb} is shown.}
	\label{fig:raa_23_dmeson_kappa_lhc_b36}
\end{figure}

\begin{figure}
	\centering
\includegraphics[width=\linewidth]{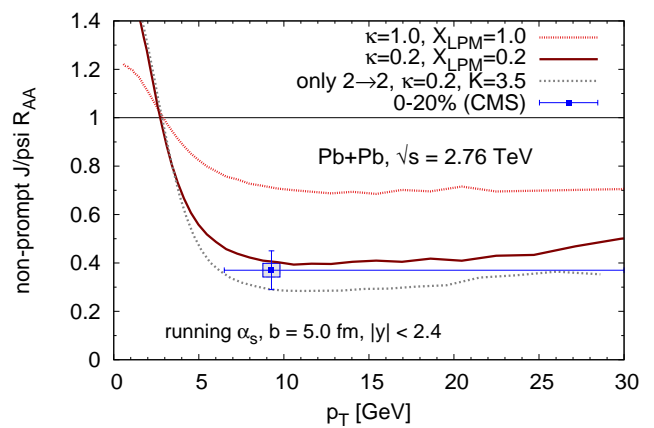}
	\caption[Non-prompt $J/\psi$ $R_{AA}$ at LHC]{$R_{AA}$ of non-prompt $J/\psi$ at LHC for $b=5.0\unit{fm}$ with data \cite{Chatrchyan:2012np} for the same configurations as in Fig.~\ref{fig:raa_23_dmeson_kappa_lhc_b36}.}
	\label{fig:raa_23_nonPromptJpsi_kappa_lhc_b50}
\end{figure}

\begin{figure}
	\centering
\includegraphics[width=\linewidth]{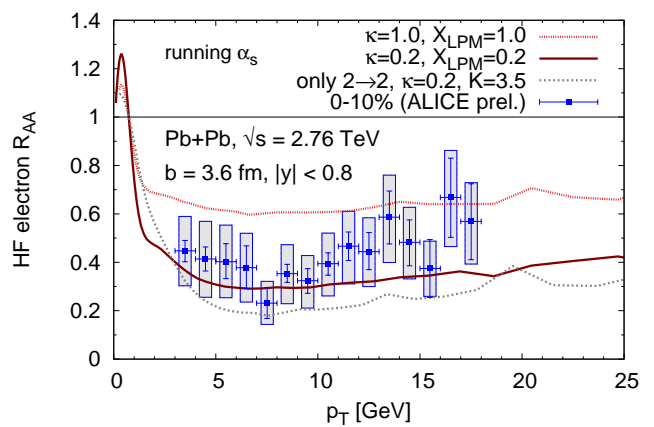}
	\caption[Heavy flavor electron $R_{AA}$ at LHC]{$R_{AA}$ of heavy flavor electrons at LHC for $b=3.6\unit{fm}$ with data \cite{delValle:2012qw} for the same configurations as in Fig.~\ref{fig:raa_23_dmeson_kappa_lhc_b36}.}
	\label{fig:raa_23_electron_kappa_lhc_b36}
\end{figure}

Figures \ref{fig:raa_23_dmeson_kappa_lhc_b36} and \ref{fig:raa_23_nonPromptJpsi_kappa_lhc_b50} depict the nuclear modification factor of $D$ mesons and non-prompt \jpsi, respectively, in Pb+Pb events at LHC.%
Again, the curves with elastic and radiative processes and $X_\text{LPM}=1$ are significantly larger than the data, but the curves with $X_\text{LPM}=0.2$ agree very well with the experimental results for both $D$ mesons and non-prompt \jpsi. This indicates that the mass hierarchy of heavy quarks is accurately described within BAMPS if the improved screening procedure is employed. 
To this end, it is not surprising that the heavy flavor electrons from $D$ and $B$ meson decays at LHC agree also with the data for $X_\text{LPM}=0.2$, as shown in Fig.~\ref{fig:raa_23_electron_kappa_lhc_b36}.

\begin{figure}
	\centering
\includegraphics[width=\linewidth]{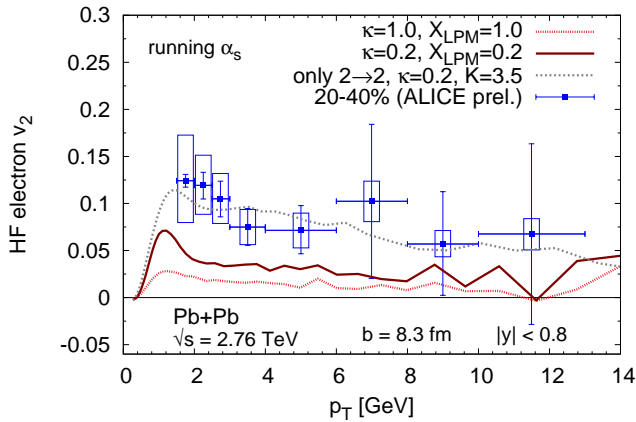}
	\caption[Heavy flavor electron $v_2$ at LHC]{Elliptic flow $v_2$ of electrons at LHC for $b = 8.3 \, {\rm fm}$ together with data \cite{Sakai:2013ata} for the same configurations as in Fig.~\ref{fig:raa_23_dmeson_kappa_lhc_b36}.}
	\label{fig:v2_23_electron_kappa_lhc_b83}
\end{figure}

\begin{figure}
	\centering
\includegraphics[width=\linewidth]{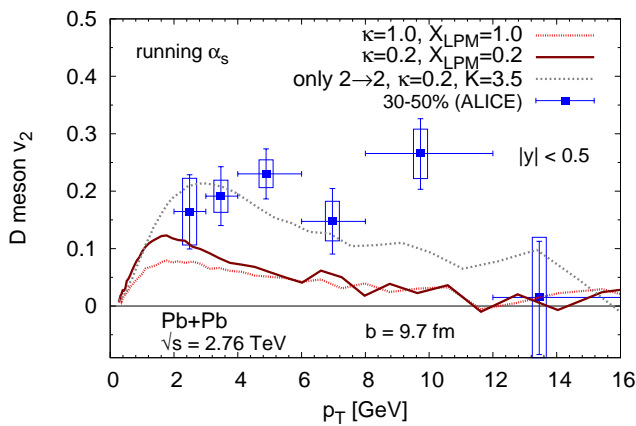}
	\caption[$D$ meson $v_2$ at LHC]{Elliptic flow $v_2$ of $D$ mesons at LHC with $b = 9.7 \, {\rm fm}$ together with data \cite{Abelev:2013lca} for the same configurations as in Fig.~\ref{fig:raa_23_dmeson_kappa_lhc_b36}.}
	\label{fig:v2_23_dmeson_kappa_lhc_b97}
\end{figure}

Figures \ref{fig:v2_23_electron_kappa_lhc_b83} and \ref{fig:v2_23_dmeson_kappa_lhc_b97} display the elliptic flow of heavy flavor electrons and $D$ mesons, respectively, in non-central events at LHC.%
Analogously to RHIC, the elliptic flow at LHC cannot be explained by elastic and radiative processes with $X_\text{LPM}=0.2$, although the \raa is described with this parameter. In contrast, the curve with only binary collisions scaled with $K=3.5$ describes both the electron and $D$ meson \raa. The explanation for the discrepancy between the two curves is the same as mentioned in the discussion of the RHIC results: the transport cross sections of the scaled binary collisions are larger than for the scenario with binary and radiative processes and $X_\text{LPM}=0.2$, although the energy loss is very similar. Hence, both have a rather similar \raa (see Figs.~\ref{fig:raa_23_dmeson_kappa_lhc_b36}, \ref{fig:raa_23_nonPromptJpsi_kappa_lhc_b50}, and~\ref{fig:raa_23_electron_kappa_lhc_b36}) but a different \vt.

A possible reason why we obtain a small elliptic flow with elastic and radiative processes might be that additional effects are missing. We neglect initial stage fluctuations, apply only Peterson fragmentation instead of coalescence at small transverse momenta, and do not take hadronic interactions into account. Furthermore, we neglect $3\rightarrow 2$ processes, which can have a sizeable contribution in the low $p_T$ region. All of these effects are expected to increase the elliptic flow, while the nuclear modification factor at large $p_T$ is not strongly affected.

Most of the open heavy flavor models in the literature that include radiative interactions cannot describe the nuclear modification factor and elliptic flow simultaneously. Both in the ASW \cite{Armesto:2005mz} and GLV (or the extended WHDG) \cite{Djordjevic:2005db,Wicks:2005gt,Buzzatti:2011vt,Buzzatti:2012pe} framework the \raa is well described (also for light partons), but the calculated \vt is significantly smaller than the data. The same is true for the results of \Refs~\cite{Cao:2011et,Cao:2012au}, where radiative processes have been implemented in a Langevin approach coupled to a $2+1$D viscous hydrodynamic model.


In contrast, the Nantes group explains the \raa and \vt data simultaneously fairly well if elastic and radiative interactions are rescaled with $K=0.7$ \cite{Gossiaux:2012ya,Nahrgang:2013saa}. This result is at first sight surprising, given that in BAMPS quite similar physics is implemented and---as we saw in this section---BAMPS cannot describe \raa and \vt simultaneously with radiative collisions. However, although the Gunion-Bertsch matrix element is also employed in \Refs~\cite{Gossiaux:2012ya,Nahrgang:2013saa}, two phenomena are treated differently compared to BAMPS. First, the emitted gluon acquires a finite mass from screening effects, whereas it is massless in BAMPS. Second, in contrast to BAMPS (see Sec.~\ref{sec:lpm} for our implementation), the LPM effect is implemented through an interpolation between single and multiple scatterings \cite{Gossiaux:2012cv} that is matched to the BDMPS result \cite{Baier:1996kr}. Both effects lead to very different properties of radiative processes in BAMPS and the Nantes model  \cite{gossiaux:private13}, although the underlying matrix element is the same. Furthermore, differences in the background medium evolution might also lead to additional differences in the heavy quark observables.

The authors of \Refs~\cite{He:2011qa} and \cite{Lang:2012yf,Lang:2012cx} find that within the resonance scattering approach the \raa and \vt can be reasonably well described simultaneously if coalescence is taken into account. The driving force behind this is the $s$ channel process, which leads to large angle scatterings and, thus, large transport cross sections.

\section{Conclusions}
\label{sec:conclusions}

In this paper we studied the interactions of heavy quarks in the partonic medium created in ultra-relativistic heavy-ion collisions at RHIC and LHC with the partonic transport description BAMPS. In addition to elastic interactions that have been previously implemented in BAMPS, we introduced in this paper also radiative heavy flavor process. To this end, we extended the recent calculation of the improved Gunion-Bertsch matrix element to finite masses and showed explicitly that the dead cone suppressions also occurs in our calculation. However, we found in the energy loss calculation of charm and light quarks in a static medium that the implementation of the LPM effect produces a second dead cone for both light and heavy particles that overshadows the original dead cone due to the heavy quark mass. Consequently, small angle radiation off both charm and light quarks is suppressed and both have a very similar radiative energy loss. 

A similar suppression of light and charm quarks is also found in full BAMPS simulations of heavy-ion collisions at LHC. In connection with the stronger suppression of gluons due to a larger color factor and mass effects in the fragmentation process of high energy partons, we found that the nuclear modification factor of $D$ mesons and charged hadrons takes very similar values. Furthermore, we found a good agreement with the experimental data of both species for standard Debye screening and the same LPM parameter.

Considering the improved Debye screening extracted from hard-thermal-loop calculations for heavy flavor processes, we find a good agreement with the experimental data for the nuclear modification factor of heavy flavor electrons at RHIC and $D$ mesons, non-prompt \jpsi, and heavy flavor electrons at LHC when taking both elastic and radiative processes into account. However, the elliptic flow of $D$ mesons and heavy flavor electrons cannot be simultaneously described. This is in contrast to previous BAMPS calculations with only elastic interactions. After multiplying the cross sections by a phenomenological factor of $K=3.5$ both observables could be simultaneously described  fairly well. The reason for this lies in a smaller transport cross section of elastic and radiative processes compared to only elastic scatterings scaled with $K=3.5$, although the energy loss of both scenarios is rather similar.

From the  observations in this paper one can conclude that radiative processes do not enhance the cross section substantially at low transverse momenta and can therefore not fully account for the build-up of the elliptic flow. It seems that the kinematics of binary interactions---either small-angle-peaked pQCD elastic processes with a $K$ factor or $s$ channel dominated resonance scatterings---are needed to get an agreement with the experimental elliptic flow data. Or in other words, heavy quarks must have a large cross section---or at least a large transport cross section as in the case of resonance scatterings---for interactions with the other medium constituents to pick up the elliptic flow of the medium. 

However, as we also showed in this work, the inclusion of radiative processes influences the results on the nuclear modification factor and elliptic flow differently---the \raa is strongly decreased but the \vt only slightly increased. Hence, it is not clear whether models that include only binary scatterings (via pQCD or resonance scattering) can still describe the nuclear modification factor and elliptic flow of heavy flavor particles simultaneously if radiative processes are taken into account.

Possible effects like initial fluctuations or coalescence have been neglected in this study. Both of them are expected to lead to an increase of the elliptic flow. Furthermore, $3\rightarrow 2$ processes for heavy quarks are not taken into account in this study, which can also enhance the elliptic flow. However, in a previous study \cite{Uphoff:2014cba} we found that with $2\rightarrow 2$ and $2 \leftrightarrow 3$ interactions the partonic elliptic flow is dominated by the elliptic flow of gluons. Light quarks have a significantly smaller flow due to a smaller color factor. If heavy quarks interact via the same mechanisms as light quarks with other medium constituents, heavy quark elliptic flow should not be larger than that of light quarks. Consequently, it is not clear how heavy particles dominated by heavy quarks and light particles dominated by gluons could have a similar flow as indicated by data when pQCD properties imply a larger color factor (and therefore larger scattering rate) for the latter. However, the picture is significantly complicated by hadronization and especially the role of gluons in it. Therefore, more detailed studies, in particular, on the hadronization mechanism are needed to draw more definite conclusions.

\section*{Acknowledgements}
We would like to thank J.\ Aichelin, P.B.\ Gossiaux, and H.\ van Hees for their constant interest in this work and  valuable discussions.
This work was supported by the Bundesministerium f\"ur Bildung und Forschung (BMBF), the NSFC, the MOST under Grants No.~11275103, No.~11335005, No.~2014CB845400, HGS-HIRe, H-QM, and the Helmholtz International Center for FAIR within the framework of the LOEWE program launched by the State of Hesse. Numerical computations have been performed at the Center for Scientific Computing (CSC).

\appendix

\setcounter{secnumdepth}{0}
\section{Appendix: Gunion Bertsch matrix element for inelastic heavy and light quark scattering}
\label{sec:gb_hq}

In this section we outline the calculation of the matrix element of the process $q+Q \arr q+Q+g$ within the Gunion-Bertsch (GB) approximation. The matrix element for light quarks within this approximation was first calculated in Ref.~\cite{Gunion:1981qs} and recently revisited in Ref.~\cite{Fochler:2013epa}. In the following we extend the latter calculation to a finite mass for one of the two quarks.

\begin{figure}
	\centering
\begin{minipage}[t]{0.49\columnwidth}
\centering
\includegraphics[width=0.8\columnwidth]{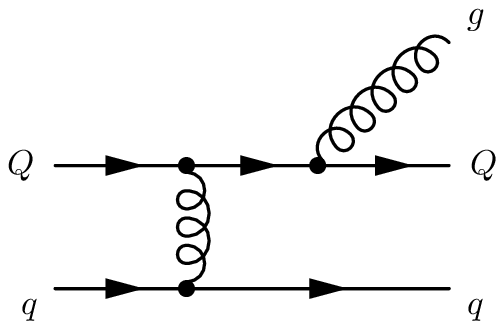}

(1)

\vspace{2em}

\includegraphics[width=0.8\columnwidth]{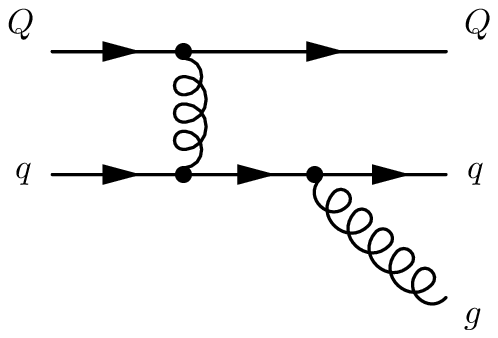}

(3)
\end{minipage}
\hfill
\begin{minipage}[t]{0.49\columnwidth}
\centering
\includegraphics[width=0.8\columnwidth]{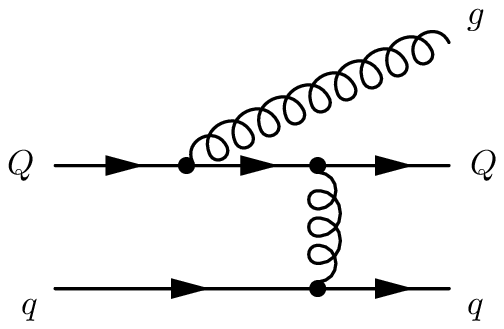}

(2)

\vspace{2em}

\includegraphics[width=0.8\columnwidth]{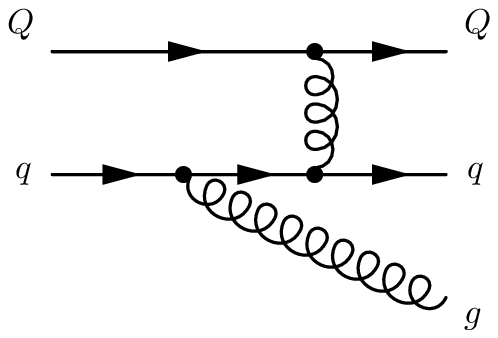}

(4)
\end{minipage}

\vspace{2em}

\includegraphics[width=0.4\columnwidth]{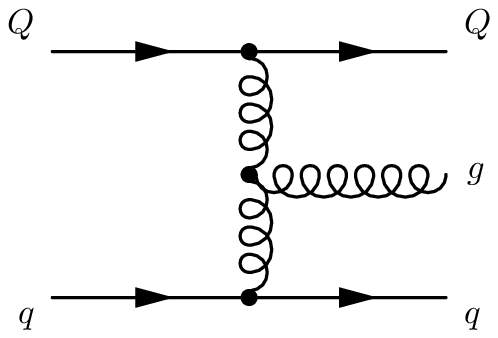}

(5)
\caption[Feynman diagrams for $qQ \arr qQg$]{Feynman diagrams for $q+Q \arr q+Q+g$.}
\label{fig:feynman_diagrams_qQ_qQg}
\end{figure}

The Feynman diagrams for the process $qQ \arr qQg$ are given in  Fig.~\ref{fig:feynman_diagrams_qQ_qQg}.
We label the four-momentum of the incoming heavy quark with $p_1$ and the incoming light quark with $p_2$. The outgoing heavy and light quarks are $p_3$ and $p_4$, respectively. The radiated gluon is denoted with $k$ and the momentum transfer of the process or, equivalently, the momentum of the internal gluon propagator is $q$.

In the following we use light cone coordinates because these are better suited to this particular problem. In the \cm frame, the momenta of the incoming particles in light cone coordinates are given by
\begin{align}
	p_1^\mu &= \l( \sqrt{s}, \frac{M^2}{\sqrt{s}}, 0, 0 \r) \breakeq
	p_2^\mu &= \l( 0, \frac{s-M^2}{\sqrt{s}}, 0, 0 \r) \ .
\end{align}
The Lorentz-invariant quantity
\begin{align}
\label{long_momentum_frac_x_org}
  x = \frac{k^+}{p_1^+}
\end{align}
characterizes the fraction of light cone momentum carried away by the radiated gluon. It can be related to the rapidity $y$ of the emitted gluon via
\begin{align}
\label{x_def_y}
x = \frac{k_\perp}{\sqrs} {\rm e}^y\ .
\end{align}

The Gunion and Bertsch matrix element is derived in the high-energy limit. This means that the radiated gluon and the momentum transfer of the process are soft,
\begin{align}
\label{gb_constraints_1}
	k_\perp &\ll \sqrt{s} \breakeq
	q_\perp &\ll \sqrt{s} \ .
\end{align}
The third approximation relates $k_\perp$, $q_\perp$, and $x$,
\begin{align}
\label{gb_constraints_2}
	x q_\perp &\ll k_\perp \ .
\end{align}
These approximations are explicitly stated by GB in their paper. However, as discussed in Ref.~\cite{Fochler:2013epa} we need in our calculation three additional approximations, that is, $x (1-x) s \gg k_{\perp}^{2}$ and all components of $q$ and $k$ being soft, $q^\mu,\, k^\mu \ll \sqrt{s}$, to end up with the same result as GB.
With the approximations~\eqref{gb_constraints_1} as well as~\eqref{gb_constraints_2} and from the constraint that all external particles are on-shell one can explicitly specify $k$ and $q$ in light cone coordinates
\begin{align}
\label{gb_k_def}
	k^\mu &= \l( x \sqrs, \frac{k_\perp^2}{x\sqrs}, {\bf k}_\perp  \r) \breakeq
	q^\mu &= \Biggl(  - q^2_\perp \frac{\sqrs}{s-M^2}, \frac{\l({\bf q}_\perp - {\bf k}_\perp\r)^2}{(1-x)\sqrs} + \frac{k_\perp^2}{x\sqrs} + \frac{x}{1-x} \frac{M^2}{\sqrs} , \Biggr. \breakeq
    & \qquad\qquad \Biggl. {\bf q}_\perp  \Biggr) \ .
\end{align}

The original computation of the GB matrix element has been carried out in light cone gauge, $A^+ = 0$. This implies that the $+$~component of the polarization vector $\epsilon(k)$ is also zero. The physical polarizations of the emitted gluon must be transverse to its momentum, $\epsilon(k) \cdot k = 0$. With these two constraints the two physical polarization vectors, $i=1,2$, are given by
\begin{align}
	\epsilon^\mu_{(i)}(k) = \l( 0, \frac{2 {\boldsymbol \epsilon}^{(i)}_\perp \cdot {\bf k}_\perp}{x \sqrs}, {\boldsymbol \epsilon}^{(i)}_\perp \r) \ ,
\end{align}
where ${\boldsymbol \epsilon}^{(1)}_\perp = (1,0)$ and ${\boldsymbol \epsilon}^{(2)}_\perp = (0,1)$ are possible choices. For brevity we will suppress the polarization index $(i)$ in the following, keeping in mind that the final polarization sum will simply amount to the replacement $\sum_{\epsilon} \left| {\bf p}\cdot{\boldsymbol \epsilon}_{\perp} \right|^{2} = {\bf p}^{2}$.

In the following we give the results of the individual diagrams.

\subsubsection{Diagram 1}
The Feynman rules give for diagram 1
\begin{align}
\label{gb_diagram1_feyn_rules}
	i \mathcal{M}_1^{qQ} &=
	\bar{u}(p_3) \, \epsilon_\mu^\star(k) \, i g \gamma^\mu  \lambda^a_{kj}\, \frac{i({\not}p_3+{\not}k+M)}{(p_3+k)^2-M^2} \, i g \gamma^\nu  \lambda^b_{ji}\, 
\breakeq &\qquad \times
u(p_1) \,
	\frac{-i g_{\nu\sigma} \delta^{bc}}{q^2} \,
	\bar{u}(p_4) \, i g \gamma^\sigma  \lambda^c_{nm}\, u(p_2) \breakeq  
  &= - i \frac{g^3}{q^2} \, \lambda^a_{kj} \lambda^b_{ji} \lambda^b_{nm}  \, \frac{1}{(p_3+k)^2-M^2} \, \epsilon_\mu^\star(k) \,
\breakeq &\qquad \times
 \Gamma^{\mu\nu}(p_{3},p_{3}+k,p_{1})   \, J_{\nu}(p_{4},p_{2})  \ ,
\end{align}
where we defined
\begin{align}
\label{gb_vertex_gamma}
 \Gamma_{rs}^{\mu\nu}(p,k,p') &= \overline{u}^{r}(p) \, \gamma^{\mu} \, ({\not} k + m) \, \gamma^{\nu} \, u^{s}(p')
\breakeq
  &= \sum_{t} J_{rt}^{\mu}(p,k) J_{ts}^{\nu}(k,p') \ ,
\end{align}
but again suppressed the spin indices in Eq.~\eqref{gb_diagram1_feyn_rules}.
With $(p_3+k)^2 = M^2 + 2 p_3 \cdot k$ and employing the eikonal approximation,
\begin{align}
\label{gb_source_eikonal}
	J^\mu(p,p+q) \simeq (2p+q)^\mu \ ,
\end{align}
for the sources the matrix element reads
\begin{align}
\label{gb_me_1_preresult}
i \mathcal{M}_1^{qQ}  &\simeq - i \frac{g^3}{q^2} \, \lambda^a_{kj} \lambda^b_{ji} \lambda^b_{nm}  \, \frac{1}{2 p_3 \cdot k} \, \epsilon_\mu^\star(k) \, (2p_1 + 2q - k)^\mu \, 
\breakeq &\qquad \times
(2p_1+q)^\nu (2p_2-q)_\nu \breakeq
	&\simeq - g \, \lambda^a_{kj} \lambda^b_{ji} \lambda^b_{nm} i M_0^{qQ} \, (1-x) \, \frac{2 \, {\boldsymbol \epsilon}_\perp \cdot {\bf k}_\perp}{k_\perp^2+x^2M^2} 	\ .
\end{align}
In the last step we inserted the matrix element for the elastic scattering without color factor
\begin{align}
\label{gb_me_0_wocolor}
	i M^{qQ}_0 = \frac{i g^2}{q^2} J_\nu(p_3,p_1) J^\nu (p_4,p_2) \ .
\end{align}
To end up at the GB result when calculating the term $(2p_1+q)^\nu (2p_2-q)_\nu$ the restriction $x (1-x) s \gg k_{\perp}^{2}$ is required in addition to the standard GB approximations \eqref{gb_constraints_1} and \eqref{gb_constraints_2}. 

\subsubsection{Diagram 2}
Diagram 2 is given by
\begin{align}
	i \mathcal{M}_2^{qQ} &=
	\bar{u}(p_3) \, i g \gamma^\nu  \lambda^b_{kj}\,  \frac{i({\not}p_1-{\not}k+M)}{(p_1-k)^2-M^2}  \, \epsilon_\mu^\star(k) \, i g \gamma^\mu  \lambda^a_{ji}\, 
\breakeq &\qquad \times
u(p_1) \,
	\frac{-i g_{\nu\sigma} \delta^{bc}}{q^2} \,
	\bar{u}(p_4) \, i g \gamma^\sigma  \lambda^c_{nm}\, u(p_2) \ .
\end{align}
With $(p_1-k)^2 = M^2 - 2 p_1 \cdot k$ and the relations \eqref{gb_vertex_gamma} and \eqref{gb_source_eikonal} the matrix element reads
\begin{align}
\label{gb_me_2_preresult}
	i \mathcal{M}_2^{qQ} &\simeq  i \frac{g^3}{q^2} \, \lambda^b_{kj} \lambda^a_{ji} \lambda^b_{nm}  \, \frac{1}{2 p_1 \cdot k} \,  \epsilon_\mu^\star(k) \, (2p_1 - k)^\mu 
\breakeq &\qquad \times
(2p_1+q-2k)^\nu (2p_2-q)_\nu
  \breakeq
	&\simeq  g \, \lambda^b_{kj} \lambda^a_{ji} \lambda^b_{nm} i M_0^{qQ} \, (1-x) \, \frac{2 \, {\boldsymbol \epsilon}_\perp \cdot {\bf k}_\perp}{k_\perp^2+x^2M^2} 	\ .
\end{align}
Again the constraint $x (1-x) s \gg k_{\perp}^{2}$ needs to be used.

\subsubsection{Diagram 3}
For diagrams 3 and 4 the kinematics are slightly different than for the other diagrams since the gluon is emitted from the lower line. Therefore, the components of the momentum transfer $q$ need to be redetermined from the on-shell conditions ($p_3^2=(p_1+q)^2=0$ and $p_4^2=(p_2-q-k)^2=0$). With the choice of keeping $k$ in the form given in Eq.~\eqref{gb_k_def}, the momentum transfer reads
\begin{equation}
q^\mu \simeq \l( -  x\sqrs, \frac{q_\perp^{2} - M^2}{\sqrs}, {\bf q}_\perp  \r) \ . 
\end{equation}
Employing the Feynman rules and performing similar simplifications as for the aforementioned diagrams give for diagram 3
\begin{align}
	i \mathcal{M}_3^{qQ} &=
  \bar{u}(p_3) \, i g \gamma^\sigma  \lambda^b_{ki}\, u(p_1) \,
	\frac{-i g_{\nu\sigma} \delta^{bc}}{q^2}  \,
	\bar{u}(p_4) \, \epsilon_\mu^\star(k) \, i g \gamma^\mu  
\breakeq &\qquad \times
\lambda^a_{nj}\, \frac{i({\not}p_4+{\not}k+M)}{(p_4+k)^2} \, i g \gamma^\nu  \lambda^c_{jm}\, u(p_2) \breakeq
  &\simeq
  - i \frac{g^3}{q^2} \, \lambda^a_{nj} \lambda^b_{jm} \lambda^b_{ki} \frac{1}{2 p_4 \cdot k}
  \epsilon_\mu^\star(k) (2p_2 - 2q - k)^\mu
\breakeq &\qquad \times
 (2p_1+q)_\nu (2p_2 - q)^\nu \breakeq
  &\simeq - g \, \lambda^a_{nj} \lambda^b_{jm} \lambda^b_{ki} i M_0^{qQ} \, (2-x) \, \frac{ {\boldsymbol \epsilon}_\perp \cdot ({\bf q}_\perp+ {\bf k}_\perp)}{s-M^2} \ .
\end{align}
In the light cone gauge $A^+$ and the high-energy limit, the contribution from this diagram is much smaller than that of diagrams 1 and 2 due to the $s$ term in the denominator. Therefore, we can neglect it in the following.

\subsubsection{Diagram 4}
For diagram 4 we get
\begin{align}
	i \mathcal{M}_4^{qQ} &=
  \bar{u}(p_3) \, i g \gamma^\sigma  \lambda^b_{ki}\, u(p_1) \,
	\frac{-i g_{\nu\sigma} \delta^{bc}}{q^2}  \,
	\bar{u}(p_4) \, i g \gamma^\nu  \lambda^c_{nj}\, 
\breakeq &\qquad \times
\frac{i({\not}p_2-{\not}k+M)}{(p_2-k)^2} \, i g \gamma^\mu  \lambda^a_{jm} \, \epsilon_\mu^\star(k)\, u(p_2) \breakeq
  &\simeq
  i \frac{g^3}{q^2} \, \lambda^b_{ki} \lambda^b_{nj} \lambda^a_{jm} \frac{1}{2 p_2 \cdot k}
  \epsilon_\mu^\star(k) (2p_2 - k)^\mu 
\breakeq &\qquad \times
(2p_1+q)_\nu (2p_2 - 2k - q)^\nu \breakeq
  &= 0 \ ,
\end{align}
since $\epsilon_\mu^\star(k) (2p_2 - k)^\mu = 0$.

\subsubsection{Diagram 5}
The last diagram is the most interesting one due to its three-gluon vertex. With help of the Feynman rules the matrix element can be written as
\begin{align}
	i \mathcal{M}_5^{qQ} &=
	\bar{u}(p_3) \, i g \gamma^\nu  \lambda^c_{ki}\, u(p_1) \, \frac{-i}{(q-k)^2} \,
	g f^{cba} 
\breakeq &\qquad \times
\l[ g_{\nu \sigma} (k-2q)^\mu + g_{\sigma \mu} (q+k)^\nu + g_{\mu \nu} (q-2k)^\nu \r] \breakeq
	 & \qquad
	 \times \epsilon_\mu^\star(k)
	\frac{-i}{q^2} \,
	\bar{u}(p_4) \, i g \gamma^\sigma  \lambda^b_{nm}\, u(p_2) \breakeq
 &= \frac{g^3}{q^2 (q-k)^2} 	f^{cba} \lambda^c_{ki} \lambda^b_{nm}\, \epsilon_\mu^\star(k) 
\breakeq &\qquad \times
 \l[ J_\nu(p_3,p_1) J^\nu (p_4,p_2) (k-2q)^\mu  \r. \breakeq
 & \qquad \qquad
 \l. + \; J_\nu(p_3,p_1) J^\mu (p_4,p_2) (q+k)^\nu  \r.
\breakeq &\qquad \qquad  \l.
+ J^\mu(p_3,p_1) J_\nu (p_4,p_2) (q-2k)^\nu \r]	 \breakeq
 &\simeq -  i g	f^{abc} \lambda^c_{ki} \lambda^b_{nm}\, iM_0^{qQ} \, (1-x) \, 
\breakeq &\qquad \times
\frac{2 \, {\boldsymbol \epsilon}_\perp \cdot ({\bf q}_\perp - {\bf k}_\perp)}{({\bf q}_\perp - {\bf k}_\perp)^2+x^2M^2}  \ .
\end{align}
Also for diagram 5 it was necessary to employ the constraint $x (1-x) s \gg k_{\perp}^{2}$ in addition to the standard GB approximations.

\subsubsection{Total matrix element}
The total matrix element is the sum of all the five diagrams where we can neglect diagram 3 and diagram 4 does not contribute. Using $([\lambda^a,\lambda^b])_{ki} = i f^{abc} \lambda^c_{ki}$ for the sum of diagrams~1 and 2 yields for the total matrix element
\begin{align}
\label{gb_me_all_diagrams}
	i \mathcal{M}_{qQ \arr qQg} &\simeq -  i g	f^{abc} \lambda^c_{ki} \lambda^b_{nm}\, iM_0^{qQ} \, (1-x) \, 2 \, {\boldsymbol \epsilon}_\perp 
\breakeq &\qquad 
\cdot \l[ \frac{ {\bf k}_\perp}{k_\perp^2+x^2M^2} +  \frac{ {\bf q}_\perp - {\bf k}_\perp}{({\bf q}_\perp - {\bf k}_\perp)^2+x^2M^2} \r] \, .
\end{align}

\bibliography{hq,text}

\end{document}